\documentclass[reprint,twocolumn,superscriptaddress,showpacs,
nofootinbib,notitlepage,floatfix]{revtex4-1}

\usepackage{graphicx}
\usepackage{latexsym,amsmath,amssymb,lmodern,float,url}
\usepackage{natbib}
\usepackage{color}
\usepackage{microtype}
\usepackage{slashed}
\usepackage{multirow}
\usepackage{bm}

\newcommand{\be}{\begin{equation}}
\newcommand{\ee}{\end{equation}}
\newcommand{\bea}{\begin{eqnarray}}
\newcommand{\eea}{\end{eqnarray}}

\newcommand{\ba}{\begin{array}}
\newcommand{\ea}{\end{array}}

\newcommand{\rjp}{R(J/\psi)}
\newcommand{\rec}{R(\eta_c)}
\newcommand{\jp}{J/\psi}
\newcommand{\bc}{B_c^+}
\newcommand{\ec}{\eta_c}
\newcommand{\lb}{\Lambda_b}
\newcommand{\lc}{\Lambda_c}
\newcommand{\bs}{B_s}
\newcommand{\ds}{D_s}
\newcommand{\tr}{\text{Tr}}
\newcommand{\gf}{\gamma_5}
\newcommand{\ga}{\gamma}
\newcommand{\tv}{\tilde{v}}

\usepackage[colorlinks=true,backref=false, linktocpage=true,
citecolor=blue,urlcolor=blue,linkcolor=blue,
pdfpagemode=UseOutlines]{hyperref}
\usepackage{cleveref}
\hypersetup{%
  bookmarksnumbered=true,
  pdftitle = {},
  pdfsubject = {},
  pdfauthor = {},
  pdfkeywords = {}
}

\newcommand{\rjpv}{0.25(3)}
\newcommand{\recv}{0.30(5)}
\newcommand{\rdv}{0.298(6)}
\newcommand{\rdstv}{0.252(14)}
\newcommand{\rdsv}{0.300(5)}
\newcommand{\rdsstv}{0.20(3)}
\newcommand{\rlcv}{0.332(10)}

\newcommand{\pjpv}{--0.47(5)}
\newcommand{\pecv}{0.33(11)}
\newcommand{\pdv}{0.325(4)}
\newcommand{\pdstv}{--0.51(5)}
\newcommand{\pdsv}{0.323(18)}
\newcommand{\pdsstv}{--0.49(5)}
\newcommand{\plcv}{--0.308(15)}

\newcommand{\fjpv}{0.46(4)}
\newcommand{\fdstv}{0.45(3)}
\newcommand{\fdsstv}{0.44(5)}

\begin{document}
\title{Precision Model-Independent Bounds from Global Analysis of
$b \to c \ell \nu$ Form Factors}
\author{Thomas D. Cohen}
\email{cohen@umd.edu}
\affiliation{Department of Physics, University of Maryland, College
Park, MD 20742, USA}
\author{Henry Lamm}
\email{hlamm@umd.edu}
\affiliation{Department of Physics, University of Maryland, College
Park, MD 20742, USA}
\author{Richard F. Lebed}
\email{Richard.Lebed@asu.edu}
\affiliation{Department of Physics, Arizona State University, Tempe,
AZ 85287, USA}
\date{August, 2019}

\begin{abstract}
We present a model-independent global analysis of hadronic form
factors for the semileptonic decays $b\rightarrow c\ell\nu$ that
exploits lattice-QCD data, dispersion relations, and heavy-quark
symmetries.  The analysis yields predictions for the relevant form
factors, within quantifiable bounds. These form factors are used to
compute the semileptonic ratios $R(H_c)$ and various decay-product
polarizations.  In particular, we find $R(D_s^*)=0.20(3)$ and
$R(J/\psi)=0.25(3)$, predictions that can be compared to results of
upcoming LHCb measurements.  In developing this treatment, we obtain
leading-order NRQCD results for the nonzero-recoil relations between
the $B_c^+ \rightarrow \{J/\psi , \eta_c \}$ form factors.
\end{abstract}

\maketitle
\section{Introduction}

While the Higgs interaction is the only source of {\it lepton
universality\/} violations within the Standard Model (SM), the
observation of neutrino masses implies that at least one form of
beyond-SM modification exists, specifically in the lepton sector.  The
factorization of QCD dynamics from electroweak interactions in the SM
allows amplitudes for semileptonic decays to be expressed as the
familiar product of hadron ($H^{\mu\nu}$) and lepton ($L^{\mu\nu}$)
tensors at leading order:
\begin{equation}\label{eq:matel}
|\mathcal{M}_{\bar{b}\rightarrow\bar{c} \, \ell^+ \nu_\ell}|^2=
\frac{L_{\mu\nu}H^{\mu\nu}}{q^2-M_W^2}+\mathcal{O}(\alpha,G_F)\,.
\end{equation}
Heavy-hadron semileptonic decay rates (both full and differential)
producing distinct lepton flavors differ only due to factors of
lepton mass that arise from kinematic and chirality-flip factors.
Such dependences can be removed in a variety of
ways~\cite{Colangelo:2018cnj,Ivanov:2016qtw,Tran:2018kuv,
Bhattacharya:2015ida,Bhattacharya:2016zcw,Jaiswal:2017rve,
Bhattacharya:2018kig,Cohen:2018vhw,Becirevic:2019tpx}.  Measurements
from BaBar, Belle, and LHCb of the ratios
$R(D^{(*)})$~\cite{Lees:2012xj,Lees:2013uzd,Huschle:2015rga,
Abdesselam:2019dgh,Sato:2016svk,Aaij:2015yra,Hirose:2016wfn,
Aaij:2017uff,Aaij:2017deq,Hirose:2017dxl} of the heavy-light meson
decays $B \! \rightarrow \!  D^{(*)}\ell\nu$, with $\ell \! = \!
\tau$ to $\ell \! = \! \mu$, exhibit a combined 3.1$\sigma$
discrepancy from the HFLAV-suggested SM values~\cite{Amhis:2016xyh},
which average~Refs.~\cite{Bigi:2016mdz,Bernlochner:2017jka,
Bigi:2017jbd,Jaiswal:2017rve}.  Recently, the LHCb collaboration has
measured $R(\jp)$~\cite{Aaij:2017tyk}, which is within
1.3$\sigma$~\cite{Cohen:2018dgz} of the SM prediction.  These results,
including lattice-determined and theoretically computed values of
$R(H_c)$, are compiled in Table~\ref{tab:models}.

\begin{table}
\caption{\label{tab:models}Existing results for $R(H_c)$ from
experiment, predictions from lattice QCD alone, and theoretical
values including additional inputs.}
\begin{center}
\begin{tabular}
{c| c c c}
\hline\hline
$H_c$ & $R_{\rm exp}$ & $R_{\rm lat}$ & $R_{\rm theory}$\\
\hline
$D$&0.340(27)(13)~\cite{Lees:2012xj,Lees:2013uzd,Huschle:2015rga,
Abdesselam:2019dgh}&0.300(8)~\cite{Aoki:2019cca,Na:2015kha,
Lattice:2015rga}&0.299(3)~\cite{Amhis:2016xyh}\\
$D^*$&0.295(11)(8)~\cite{Lees:2012xj,Lees:2013uzd,Huschle:2015rga,
Sato:2016svk,Aaij:2015yra,Hirose:2016wfn,Aaij:2017uff,Aaij:2017deq,
Hirose:2017dxl,Abdesselam:2019dgh}&--&
0.258(5)~\cite{Amhis:2016xyh}\\
$\ds$&--&0.2987(46)~\cite{McLean:2019qcx}&--\\
$\lc$&--&0.3328(74)(70)~\cite{Detmold:2015aaa}&
0.324(4)\cite{Bernlochner:2018kxh}\\
$\jp$&0.71(17)(18)~\cite{Aaij:2017tyk}&
[0.20,0.39]~\cite{Cohen:2018dgz}&--\\
$\ec$&--&0.30(4)~\cite{Berns:2018vpl,Murphy:2018sqg}&--\\
\hline
\end{tabular}
\end{center}
\end{table}

In the future, it would be useful to consider other semileptonic
decays.  Run III of LHCb may open the opportunity to measure
$R(D_s^{(*)})$~\cite{Cerri:2018ypt}.  A determination of $R(\ec)$
would be exciting.  However, $\rec$ is substantially harder to measure
than $\rjp$ for a few reasons, foremost of which is the absence of a
clean $\ec$ decay process with a substantial branching fraction
(analogous to $\jp\rightarrow\mu^+\mu^-$) for reconstructing the
$\ec$; this leads to large backgrounds.  Additionally, the transition
from excited charmonium states to $\ec$ is poorly understood, which
further complicates the extraction of signals~\cite{BHHJ}.

In order to fully leverage all of these experimental results, it is
necessary to have  rigorous predictions from the SM for
all of these ratios. Even setting aside the very interesting issue of
lepton universality, determining hadronic form factors is important in
its own right, as each such function represents a wealth of information 
about nonperturbative QCD.  Form factors are not completely
unconstrained, however.  They must satisfy well-known
model-independent constraints that follow from bedrock principles of
quantum field theory, specifically unitarity and the complex
analyticity of their Green's functions as functions of momentum
variables at all values, except when a resonance, particle-creation
threshold, or other special kinematic configuration is realized.  In
the case of semileptonic decays, the form factors can be parametrized
as a product of known functions representing resonant poles and other
nonanalytic structures in the corresponding Green's function, times a
Taylor series in a conformal variable that tracks the momentum 
transfer;  the Taylor coefficients are
constrained in magnitude by unitarity.  This is the {\it BGL
parametrization\/}~\cite{Boyd:1995cf,Boyd:1997kz} (see
Ref.~\cite{Grinstein:2015wqa} for a brief review of its historical
antecedents).

The constraint of unitarity in model-independent approaches such as
the BGL parametrization has historically been underutilized, because
fits to experiment typically consider only a single exclusive process
({\it e.g.}, $B \! \to D^* \! \ell \nu$).  However, each exclusive
channel appearing in the two-point Green's function for $b \! \to \!
c$ currents positively contributes to the unitarity bound,
and therefore a simultaneous fit including multiple processes
provides stronger constraints on each of the individual
processes~\cite{Boyd:1997kz,Bigi:2017jbd,Jaiswal:2017rve}.

The purpose of this work is to perform a global analysis within the
BGL parametrization of lattice-QCD data for seven exclusive hadronic
$b\rightarrow c$ processes: $B \! \to \! \{D,D^*\}$, $B_s \! \to \!
\{D_s,D_s^{*}\}$, $B_c \! \to \! \{ \jp, \ec \}$, and $\Lambda_b \!
\to \! \Lambda_c$.  We obtain the corresponding transition form
factors.  These form factors are computed directly from the SM and
obtained within quantifiable bounds.  Thus, they can be used to make
reliable predictions from the SM that can directly confront
experiment.  We use them to compute three types of observables: the
semileptonic decay ratios $R(H_c)$, the $\tau$ polarization
$P_{\tau}(H_c)$, and the vector-meson longitudinal polarization
fraction $F^{H_C}_L$.

The analysis presented here is essentially model independent.  We do
not use {\it ad hoc\/} model assumptions about the physical
mechanisms  dominating the  form factor, beyond accepting the SM to
identify allowed functional forms of the form factors.   Rather, we
use as input  {\it ab initio\/}  Monte Carlo calculations of QCD from
the lattice as our principal input.  Such data is quite limited: not
all of the relevant form factors have been computed, and those that
have are computed at a limited number of  momentum-transfer values.
The BGL parametrization using the unitarity bound allows  us to
extend our knowledge of the form factors to other momentum transfers,
and to do so in such a way that the errors can be quantified.  We can gain
information about form factors that have not been directly computed
on the lattice from those that have by exploiting emergent symmetries
of QCD that become valid as the quark masses become large.

There are, of course, errors associated with truncating the series in
BGL parametrization and truncating the expansion in the inverse
heavy-quark mass.  Fortunately, one has {\it a priori\/} estimates of
their size, which allows for reliable  SM predictions without
invoking additional model dependence.  However, some judgment is
required in estimating their sizes quantitatively.   We have
therefore made very conservative estimates in order to ensure that
our predicted bounds  are reliable enough so that the experimental
measurements provide meaningful tests of the SM.

The lattice data that provides the input for the analysis has both
statistical and systematic  errors.  The statistical errors can be
easily incorporated into our bounds.  In some cases the major
systematic errors have been well-explored and estimated reliably, and
can also be incorporated in a straightforward way.  However, in some
cases the only available lattice calculations do not provide
estimates for some of the major systematic errors.  In these cases,
we add a systematic error in by hand, and do so in  a very
conservative manner, by assuming larger-than-realistic errors.

Thus, while the analysis is model independent, it does involve some
{\it ad hoc\/} judgment in the assignment of systematic errors.  As
this was done quite conservatively, the principal effect is to make
the error bounds for our final results larger than they otherwise
would be.  These effects can, of course, be mitigated by the constant
improvement in the treatment of errors in lattice simulations.

We begin in Sec.~\ref{sec:sm} with a discussion of the $V \! - \! A$
weak-interaction structure of the SM responsible for semileptonic
decays and the form factors under investigation.  In
Sec.~\ref{sec:hqss} we explain how heavy-quark symmetries can be used
to obtain relations between the form factors of heavy-light systems
and heavy-heavy meson systems.  The lattice results used in this work
are discussed in Sec.~\ref{sec:lat}\@.  Section~\ref{sec:da} presents
the dispersive-analysis framework utilized to constrain the form
factors as functions of momentum transfer.  The results of our
analysis are presented in Sec.~\ref{sec:results}, and we conclude in
Sec.~\ref{sec:con}.

\section{Structure of $\langle H_c|(V-A)^\mu| H_b
\rangle$}\label{sec:sm}

Since the first-principles calculation of the leptonic tensor
$L^{\mu\nu}$ in Eq.~(\ref{eq:matel}) is straightforward in the SM, the
computation of semileptonic decay rates reduces to parametrizing
exclusive components $\langle H_c|(V-A)^\mu|H_b\rangle$ of the
hadronic tensor in terms of transition form factors.  The tensor
structure is expressed in terms of the hadron momenta, $P^\mu$ for
$H_b$ (with mass $M \! \equiv \!  M_{H_b}$) and $p^\mu$ for $H_c$
(with mass $m \! \equiv \!  M_{H_c}$), and additionally the
polarization $\epsilon^\mu$ of the $H_c$ if it is a vector meson, or
heavy-quark spinors $u_{b,c}$ if $H_{b,c}$ are baryons.  The only
functional dependence of the form factors arises through
the squared momentum transfer to the leptons, $t \! = \! q^2 \! \equiv
\! (P-p)^2$.  The various cases of phenomenological interest are now
outlined.

\subsection{$B\rightarrow D, \ B_s\rightarrow D_s, \ \bc\rightarrow
\ec$}

If both $H_{b,c}$ are pseudoscalar mesons, then only two independent
Lorentz structures, and hence two independent form factors, are
possible:
\begin{eqnarray}
\lefteqn{\langle H_c(p)|(V-A)^\mu|H_b(P)\rangle} & & \nonumber \\
& = & f_+(t) (P+p)^\mu+f_- (t) (P-p)^\mu \, .
\label{eq:hadme}
\end{eqnarray}
Indeed, the parity invariance of strong interactions precludes the
current $A^\mu$ from providing a nonzero contribution to
Eq.~(\ref{eq:hadme}).  In this work, we exchange $f_-(t)$ for the
combination
\begin{equation}
\label{eq:f0}
 f_0(t) \equiv (M^2-m^2)f_+(t) +tf_-(t) \, .
\end{equation}
With this definition, one sees that $(M^2-m^2)f_+(0)=f_0(0)$, a
constraint upon two otherwise independent form factors that we will
impose when fitting the functions.  This normalization of $f_0$
differs by a mass-dependent prefactor from that used in
lattice-QCD
calculations~\cite{Colquhoun:2016osw,ALE,Aoki:2019cca,Na:2015kha,
Lattice:2015rga,McLean:2019qcx}:
\begin{equation}
\label{eq:convert}
 f_0(t)=(M^2-m^2)f_0^{\rm lat}(t) \, .
\end{equation}

The differential decay rate for this semileptonic decay process is
\begin{align}
\label{eq:difcof}
 \frac{d\Gamma}{dt}=\frac{G_F^2|V_{cb}|^2}{192\pi^3M^3}
\frac{k}{t^{5/2}}(t-m_\ell^2)^2[4&k^2t(2t+m_\ell^2)|f_+|^2\nonumber\\
&+3m_\ell^2|f_0|^2]\, .
\end{align}
where, in terms of the spatial momentum $\bm{p}$ of $H_c$ in the $H_b$
rest frame,
\begin{equation} \label{eq:kdef}
k \equiv M \sqrt{\frac{\bm{p}^2}{t}}
= \sqrt{\frac{(t_+-t)(t_--t)}{4t}} \, ,
\end{equation}
in which we have, in turn, introduced two important kinematic values,
$t_{\pm} \equiv (M\pm m)^2$.

\subsection{$B\rightarrow D^*, \ \bs\rightarrow \ds^*, \ \bc
\rightarrow \jp$}

Transition form factors of a pseudoscalar meson $H_b$ to a vector
meson $H_c$ have been parametrized in a variety of ways in
the literature.  Here, we begin with a set~\cite{Wirbel:1985ji} of
vector [$V(t)$] and axial-vector [$A_i(t)$] form factors frequently
used in lattice-QCD and model calculations:
\begin{widetext} 
\begin{align}\label{eq:hadme2}
 \langle H_c(p,\epsilon)|(V-A)^\mu|H_b(P)\rangle=&
\frac{2i\epsilon^{\mu\nu\rho\sigma}}{M+m}
\epsilon^{*}_{\nu}p_{\rho}P_{\sigma}V(t)-(M+m)
\epsilon^{*\mu}A_1(t)\nonumber\\
 &+\frac{\epsilon^{*}\cdot q}{M+m}(P+p)^\mu A_2(t)
+2m\frac{\epsilon^{*}\cdot q}{q^2}q^{\mu}A_3(t)
-2m\frac{\epsilon^{*}\cdot q}{q^2}q^{\mu}A_0(t) \, ,
\end{align}
\end{widetext}
where $q^\mu \! \equiv \! (P-p)^\mu$.
Only four of these five form factors are independent: 
demanding that only $A_0(t)$ couples to timelike virtual $W$
polarizations ($\propto q^\mu$) requires
\begin{equation}\label{eq:a3}
 A_3(t)=\frac{M+m}{2m}A_1(t)-\frac{M-m}{2m}A_2(t) \, .
\end{equation}
Requiring the cancellation of $1/q^2$ terms in Eq.~(\ref{eq:hadme2})
as $q^2 \! = \! t \! \to \! 0$ imposes the additional constraint
$A_3(0) \! = \! A_0(0)$.

A different decomposition, in which the virtual $W$ and vector meson
$H_c$ are described by their helicity states, turns out to be more
useful for the dispersive analysis.  Here, one exchanges the form
factors $V, A_0, A_1, A_2$ for the set $g, f, \mathcal{F}_1,
\mathcal{F}_2$.  They are related by
\begin{eqnarray}
 g&=&\frac{2}{M+m} V \, , \nonumber\\
 f&=&(M+m)A_1 \, , \nonumber\\
 \mathcal{F}_1&=&\frac{1}{m}\left[-\frac{2k^2 t}{M+m} A_2
-\frac{1}{2}(t-M^2+m^2)(M+m)A_1\right] \, , \nonumber\\
 \mathcal{F}_2&=&2A_0 \, . \label{eq:FFrelns}
\end{eqnarray}
 $\mathcal{F}_{1,2}$, are proportional to
the conventionally defined~\cite{Richman:1995wm} helicity amplitudes
$H_{0,t}$, respectively, while the other two helicity amplitudes
$H_\pm$ are linear combinations of $V$ and $A$ form factors, $H_\pm
(t) \! = \! f(t) \mp k \sqrt{t} g(t)$, where $k$ is defined in
Eq.~(\ref{eq:kdef}).

At  $t \! = \! t_-$, the middle two expressions of
Eqs.~(\ref{eq:FFrelns}) reduce to an additional constraint,
$\mathcal{F}_1(t_-)=(M-m)f(t_-)$.  In this basis, the previously
noted constraint $A_3(0) \! = \! A_0(0)$ becomes
$\mathcal{F}_1(0)= \frac 1 2 (M^2-m^2)\mathcal{F}_2(0)$.  The
differential decay rate for the semileptonic decay in this basis
reads
\begin{align}
\label{eq:difcofvec}
 \frac{d\Gamma}{dt}=&\frac{G_F^2|V_{cb}|^2}{192\pi^3M^3}
\frac{k}{t^{5/2}}\left(t-m_\ell^2\right)^2\nonumber\\&
\times \left\{ \left(2t+m_\ell^2\right)\left[2t|f|^2+
|\mathcal{F}_1|^2+2k^2t^2|g|^2\right] \right.
 \nonumber\\& \left. \phantom{xxxx}+3m_\ell^2k^2t|\mathcal{F}_2|^2
\right\} \, .
\end{align}
\newline
\subsection{$\lb\rightarrow \lc$}

In the case of heavy-baryon transitions, the states of the
spin-$\frac 1 2$ baryons are represented by spinors $u_{b,c}$.  Here,
there are two form factors for both the vector and axial-vector
currents:
\begin{align}
 \langle \lc(p)|V^\mu|\lb(P)\rangle& \! = \! \bar{u}_c(p)\left[
 F_1\gamma^\mu
 +F_2v^\mu \! + \! F_3 \, v'^{\mu}\right]u_b(P) \, , \nonumber\\
 \langle \lc(p)|A^\mu|\lb(P)\rangle& \! = \! \bar{u}_c(p)\left[
 G_1\gamma^\mu \!
 + \! G_2v^\mu \! + \! G_3v'^{\mu}\right] \! \ga_5u_b(P) \, ,
 \nonumber \\
\end{align}
where the kinematical variables relevant to the heavy-quark limit
(see Sec.~\ref{sec:hqss}) are the baryon 4-velocities, $v^\mu \!
\equiv \! P^\mu/M_{\lb}$ and $v^{\prime \mu} \! \equiv \! p^\mu
/M_{\lc}$.  The differential decay rate is then
\begin{widetext}
 \begin{align}
 \frac{d\Gamma}{dt}=\frac{G_F^2|V_{cb}|^2}{192\pi^3M^3}
\frac{k}{t^{5/2}}\left(t-m_\ell^2\right)^2 \big\{ &
(t_--t)(2t+m_\ell^2) [2t|F_1|^2 +|H_V|^2]+3m_\ell^2(t_+-t)|F_0|^2
\nonumber\\
+&(t_+-t)(2t+m_\ell^2)[2t|G_1|^2 \! +|H_A|^2]+3m_\ell^2(t_--t)|G_0|^2
\big\} \, ,
\end{align}
\end{widetext}
where the form factors in helicity basis read
\begin{align}
H_V= \, &(M+m)F_1+\frac{1}{2}(t_+-t)\left(\frac{F_2}{M}+\frac{F_3}{m}
\right) \, , \nonumber\\
H_A= \, &(M-m)G_1-\frac{1}{2}(t_--t)\left(\frac{G_2}{M}+\frac{G_3}{m}
\right) \, , \nonumber\\
F_0= \, &(M-m)F_1+\frac{1}{2M}(t+M^2-m^2)F_2\nonumber\, , \\&
\phantom{xxxxxxxxx}-\frac{1}{2m}(t-M^2+m^2)F_3\nonumber\, , \\
G_0= \, &(M+m)G_1-\frac{1}{2M}(t+M^2-m^2)G_2\nonumber\, , \\&
\phantom{xxxxxxxxx}+\frac{1}{2m}(t-M^2+m^2)G_3 \, . \label{eq:BaryFF}
\end{align}
As in the meson case, these form factors satisfy exact constraints at
special kinematic points.  Specifically, $H_A(t_-) \! = \! (M-m)
G_1(t_-)$, $(M+m) F_0(0) \! = \! (M-m) H_V(0)$, and $(M-m) G_0(0) \!
= \! (M+m) H_A(0)$.  This basis differs from that used in the
lattice-QCD calculations of Ref.~\cite{Detmold:2015aaa} only by
mass-dependent prefactors:
\begin{align}
F_0&=(M-m)f_0^{\rm lat} \, , \nonumber\\
H_V&=(M+m)f_+^{\rm lat} \, , \nonumber\\
F_1&=f_\perp^{\rm lat}  \, , \nonumber\\
G_0&=(M+m)g_0^{\rm lat} \, , \nonumber\\
H_A&=(M-m)g_+^{\rm lat} \, , \nonumber\\
G_1&=g_\perp^{\rm lat}  \, .
\end{align}

\section{Heavy-Quark Symmetries}\label{sec:hqss}

 The physics of heavy-light hadrons ($Q \bar q$ or $Q \bar q \bar q^\prime)$  
 is simplified by the emergence of
additional symmetries in the limit $m_Q \! \to \! \infty$.  Operators
distinguishing between heavy quarks of different spin orientation and flavor are
suppressed by $1/m_Q$, and produce a vanishingly small effect upon
physical amplitudes in the heavy-quark limit.  All transition form
factors between two hadrons with a single heavy quark and the same
light-quark content are proportional to a single, universal
Isgur-Wise function, $\xi(w)$~\cite{Isgur:1989vq,Isgur:1989ed} for
mesons or $\zeta(w)$ for baryons~\cite{Mannel:1990vg}.  They are
naturally expressed $w$, which is the dot
product of the initial and final heavy-light hadron 4-velocities,
$v^\mu \! \equiv \! P^\mu / M$ and $v'^\mu \! \equiv \! p^\mu / m$,
respectively, and fully contains the information about $t$:
\begin{equation} \label{eq:wdef}
w \equiv v \cdot v' = \gamma_m = \frac{E_m}{m} =
\frac{M^2 + m^2 - t}{2Mm} \, .
\end{equation}
The zero-recoil point, where the final hadron $m$ is created at rest
in the rest frame of the initial hadron $M$, satisfies $t \! = \! t_-
\! \equiv \! (M \! - \! m)^2$, corresponding to $w \! = \! 1$.  From
the middle expressions of Eq.~(\ref{eq:wdef}),  one notes that $w$ is
the Lorentz factor $\gamma_m$ of $m$ in the $M$ rest
frame.  The maximum value of $w$ in a given semileptonic process
occurs when the momentum transfer $t$ through the virtual $W$ to the
lepton pair---the total energy-squared of the leptons in their rest
frame---assumes its smallest possible value, $t \! = \!  m_\ell^2$.

Heavy-quark symmetry encodes a physical picture 
in which a heavy-light hadron is described by a nearly static
color-fundamental source with spin-independent interactions
(the heavy quark $Q$), to which the matter associated with light degrees
 of freedom  (light-quarks and gluons)  is bound.
 In weak decays with $Q\rightarrow Q^\prime$, the
zero-recoil (Isgur-Wise) point corresponds to a situation in which $Q$
spontaneously transforms to $Q^\prime$ at rest, but the decay
otherwise leaves the light degrees of freedom undisturbed.  The
overlap between the initial and final light-quark wave functions is
complete, so that $\xi(1) \! = \! 1$ or $\zeta(1) \! = \! 1$ at the
zero-recoil (Isgur-Wise) point.  Thus, in the heavy quark limit one 
obtains an absolute normalization for the form factors; in the meson
case~\cite{Isgur:1989vq,Isgur:1989ed,Boyd:1997kz}, all the form
factors are proportional to $\xi(w)$:
\begin{align}
\label{eq:iw}
 f_+&=\frac{1}{2}\mathcal{F}_2=\frac{1+r}{2\sqrt{r}}\xi \, ,
\nonumber \\
 f_0&=\mathcal{F}_1=M^2\sqrt{r}(1-r)(1+w)\xi \, , \nonumber\\
 g&=\frac{1}{M\sqrt{r}}\xi \, , \nonumber\\
 f&=M\sqrt{r}(1+w) \xi \, ,
\end{align}
while in the baryon case~\cite{Mannel:1990vg}, all the form factors
are proportional to $\zeta(w)$:
\begin{align}
\label{eq:iwbary}
F_0&=H_A=M(1-r)\zeta \, , \nonumber \\
F_1& = G_1 = \zeta \, , \nonumber \\
H_V&=G_0 = M(1+r)\zeta \, ,
\end{align}
where $r \! \equiv \! m/M$.  All of these results are corrected by
effects of $\mathcal{O}(\Lambda_{\rm QCD}/m_{Q^\prime})$.

Due to the lack of lattice data for $g,\mathcal{F}_1,\mathcal{F}_2$,
we use the relations of Eq.~(\ref{eq:iw}) for $f_+,f_0,$ and $f$ to
obtain $\xi_{(s)}(w)$ for each of the $B_{(s)}\rightarrow D_{(s)}$
processes from the existing lattice data.  To establish a first
approximation for an allowed region, we parametrize $\xi_{(s)}(w)$ by
\begin{equation}
 \xi_{(s)}(w)=\xi_{(s)}(1)-\rho^2(w-1)+\frac{1}{2}\sigma^2(w-1)^2.
\end{equation}
In our analysis we have included an additional systematic error of
20\% to account for violations of Isgur-Wise scaling.  We sample
three synthetic points from $\xi_{(s)}(w)$ for each form factor.  For
$B\rightarrow D$, the synthetic points are restricted to the same
range $w<1.16$ for $B\rightarrow D$ as the lattice data.  For
$\bs\rightarrow\ds$, where lattice results have been computed in the
full $w$ range, we restrict the synthetic points to the
near-zero-recoil range of $w<1.04$.

In decays of the types $\bc \! \rightarrow \! \jp(\ec) \, \ell^+
\nu_\ell$,  the spectator $c$ quark can no longer be
considered light (and indeed is the same species as the final heavy
quark).  These cases are more complicated; 
the heavy-quark limit differs from the heavy-light case in two important
ways~\cite{Jenkins:1992nb}.  First, the
heavy-quark kinetic-energy operators for $\bar b$ and $\bar c$ quarks,
while both scaling as $1/m_Q^{(\prime)}$, differ for the two flavors
(thus breaking the heavy-quark flavor symmetry), but still provide
leading-order corrections to the dynamics of the state due to the
presence of the heavy spectator $c$: {\it e.g.}, the Bohr radii of
$B_c$ and $\jp (\ec)$ are significantly different.  Second, the
spectator $c$ receives a momentum transfer due to the transition
$\bar{b} \! \to \! \bar{c}$ of the same order as the momentum imparted
to the $\bar{c}$.  Thus, the heavy-flavor symmetry due to the
replacement of $\bar b$ with $\bar c$ does not leave the spectator
degrees of freedom invariant, meaning that one cannot obtain a
normalization of the form factors at the zero-recoil point based
purely upon symmetry.

Even though the heavy-flavor symmetry obtained from replacing $\bar b$
with $\bar c$ is lost, the $\bar{b}$ and $\bar{c}$ quarks retain
separate heavy-quark spin symmetries, as does the heavy spectator $c$.
In addition, since the valence quarks are heavy  
these systems are better described using nonrelativistic dynamics
than are heavy-light systems.  Indeed, $w_{\rm max} \approx 1.3$ for
$\bc \! \rightarrow \! \jp(\ec)$, a sufficiently modest value that
suggests information obtained near the zero-recoil point remains
phenomenologically useful.  The six meson form factors of
Eqs.~(\ref{eq:hadme}) and (\ref{eq:FFrelns}) are related by the spin
symmetries to a single, universal function that
Ref.~\cite{Jenkins:1992nb} calls $\Delta$, and
Ref.~\cite{Kiselev:1999sc} calls $h$.  However, as emphasized in
Ref.~\cite{Jenkins:1992nb}, the form factors only approach $\Delta(w)$
near the zero-recoil point, and its normalization there is not fixed
by symmetry to assume a special value, like $\xi(1) \! = \! 1$.

A central feature of Ref.~\cite{Jenkins:1992nb} is the use of the
trace formalism of Ref.~\cite{Falk:1990yz} to compute the relative
normalization between the six meson form factors ({\it i.e.}, to
obtain the correct multiple of $\Delta$ for each tensor structure)
near the zero-recoil point.  To be specific, ``near'' in this sense
means kinematic configurations in which the spatial momentum transfer
to the spectator $q$ is no larger than its mass $m_q$.  This
calculation was generalized in
Refs.~\cite{Kiselev:1999sc,Kiselev:2002vz} using NRQCD to consider a
small-recoil limit ($w \! \to \! 1$) in which the four-velocities of
$\bar b$ and $\bar c$ are nevertheless unequal ({\it i.e.}, the
spectator receives a momentum transfer at leading order in NRQCD).
These relations were used in
Refs.~\cite{Cohen:2018dgz,Berns:2018vpl,Murphy:2018sqg} to constrain
the $\bc \! \rightarrow \! \jp \, (\ec)$ form factors at zero recoil.

In this work, we extend the relations
of Refs.~\cite{Kiselev:1999sc,Kiselev:2002vz} by deriving the
leading-order NRQCD relations between the form factors and $\Delta$
at non-zero recoil.  While these relations are expected to receive
large corrections away from $w=1$, we use them to construct ratios of
derivatives of form factors at $w=1$.  To proceed, we start with the
trace formalism of Ref.~\cite{Falk:1990yz}:
\begin{widetext}
\begin{align}
 \langle \{\jp,\ec\}|\bar{b}\Gamma c|\bc\rangle=-\sqrt{Mm}\tr\left[
 \frac{1+\slashed{v}_{cs}}{2}\bc\gf\frac{1-\slashed{v}_b}{2}\Gamma
 \frac{1-\slashed{v}_{c}}{2}\left(\jp^{\dag\mu}\ga_{\mu}+\ec^\dag\gf
 \right)\right]\Delta \, ,
\end{align}
\end{widetext}
where $v_{cs},v_{b},v_c$ are the velocities of the spectator $c$
quark, decaying $b$ quark, and final-state $c$ quark respectively.
These velocities are related in the heavy-quark limit to those of the
mesons by $v_b\rightarrow\tv_1$, $v_c\rightarrow\tv_2$, and $v_{cs}
\rightarrow -\tv_3=\frac{1}{2}(v_1+v_2)$, where
\begin{align}
 \tv_1^\mu=v_1^\mu+\theta\left(v_1-v_2\right)^\mu \, , \nonumber\\
 \tv_2^\mu=v_2^\mu+\omega\left(v_2-v_1\right)^\mu \, ,
\end{align}
with $\theta \! \equiv \! \frac{m_3}{2m_1}$ and $\omega \! \equiv \!
\frac{m_3}{2m_2}$.  One can then use the trace formalism to obtain the
$w$-dependent generalizations of the constants defined in
Refs.~\cite{Kiselev:1999sc,Kiselev:2002vz}.  Starting with the tensor
definitions
\begin{eqnarray}
\langle \ec (v_2) | V^\mu | B_c^+ (v_1) \rangle & \equiv &
\sqrt{m_{\ec} m_{B_c}} \left( c^P_1 v^\mu_1 + c^P_2 v^\mu_2 \right)
\Delta , \nonumber \\
\langle J/ \! \psi (v_2) | V^\mu | B_c (v_1) \rangle & \equiv &
\sqrt{m_{\jp} m_{B_c}} i c_V \varepsilon^{\mu \nu \alpha \beta}
\epsilon^{* \mu} v_{1 \alpha} v_{2 \beta} \Delta , \nonumber \\
\langle J/ \! \psi (v_2) | A^\mu | B_c (v_1) \rangle & \equiv &
\sqrt{m_{\jp} m_{B_c}} \left[  c_\epsilon \epsilon^{*\mu} + c_1 (
\epsilon^* \cdot v_1 ) v^\mu_1 \right. \nonumber \\ & & \left.
\hspace{5.5em}+ c_2 ( \epsilon^* \cdot v_2 ) \right] \Delta ,
\end{eqnarray}
where
\begin{align}
 c^P_1&=1+\theta-\omega-\frac{\omega}{2}(1+\theta)(w-1) \, ,
\nonumber\\
 c^P_2&=1-\theta+\omega-\frac{\theta}{2}(1+\omega)(w-1) \, ,
\nonumber\\
 c_V&=-1-\theta-\omega \, , \nonumber\\
 c_\epsilon&=2-\frac{\omega+\theta+2\theta\omega-2}{2}(w-1) \, ,
\nonumber\\
 c_1&=\frac{(3+2\theta)\omega}{2} \, , \nonumber\\
 c_2&=-1 -\frac{\omega}{2} -\theta(1+\omega) \, ,
\end{align}
we construct the Isgur-Wise-like relations for the heavy-heavy
systems:
\begin{align}
\label{eq:hhdelta}
 f_+&=\sqrt{r} \, \frac{c^P_1+c^P_2r^{-1}}{2}\Delta \, , \nonumber\\
 f_0&=M^2\sqrt{r}[(1-wr)c^P_1 + (w-r)c^P_2 ]\Delta
\, , \nonumber\\
 g&=-\frac{c_V}{M\sqrt{r}}\Delta \, , \nonumber\\
 f&=M\sqrt{r}c_\epsilon\Delta \, , \nonumber\\
 \mathcal{F}_1&=M^2\sqrt{r}[(w-r) c_\epsilon +(r c_1+c_2)(w^2-1)]
 \Delta \, , \nonumber\\
 \mathcal{F}_2&=\frac{c_\epsilon+ (1-wr)c_1 + (w-r)c_2} {\sqrt{r}}
  \Delta \, .
\end{align}

These relations reproduce the standard Isgur-Wise
results~\cite{Isgur:1989vq,Isgur:1989ed,Boyd:1997kz} of
Eq.~(\ref{eq:iw}) in the limit $\theta,\omega \! \rightarrow \! 0$, and
they reduce to the relations of Refs.~\cite{Kiselev:1999sc,
Kiselev:2002vz} when $w \! \to \! 1$.  Terms that break these
relations should be $\mathcal{O}(m_c/m_b, \, \Lambda_{\rm QCD}/m_c)
\approx30\%$, and, in our analysis we conservatively allow for up to 50\% violations.
We use these relations in our analysis to fix both the relative
normalization between form factors and their slopes at zero recoil.
These results are obtained by constructing ratios from
Eqs.~(\ref{eq:hhdelta}) after solving for $\Delta(w=1)$ and
$d\Delta/dw|_{w=1}$.

\section{Lattice QCD Results}\label{sec:lat}

This work we uses the existing lattice-QCD
results for $b\rightarrow c$ form factors as input to our global 
analysis.  These results have been
produced by a number of different groups, and the determinations of
the various form factors have been performed at varying numbers of
momentum-transfer values $t$ and with varying treatments of
uncertainties.  In this section we summarize the lattice results
used in our analysis.

The best current results are those for $B \! \rightarrow \! D$ form
factors.  The form factors $f_+$ and $f_0$ have been computed by two
groups~\cite{Na:2015kha,Lattice:2015rga}, including a complete
treatment of all sources of error.  We use the
results of~\cite{Lattice:2015rga} alone in the final results, having
found that the larger uncertainties of \cite{Na:2015kha} mean that
they provide no significant additional constraint.

For the case of $B_s\rightarrow D_s$, a single group has produced
results for $f_+$ and $f_0$ at non-zero recoil with a complete error
treatment~\cite{McLean:2019qcx}.  The baryonic process
$\lb\rightarrow\lc$ has been computed in Ref.~\cite{Detmold:2015aaa}
on only one lattice volume, but their results include a 1.5\%
systematic uncertainty for finite-volume effects, and given a quantified error
estimate,  we can include this lattice data in our analysis.

The heavy-heavy process $\bc \! \rightarrow \! \ec$ has also only been
computed by one group~\cite{Colquhoun:2016osw,*ALE}, with an
incomplete treatment of errors.  It was computed on a single lattice
volume, so finite-volume effects are potentially worrying.  To account for
possible large finite-volume effects, in the analysis we included
an additional 20\% systematic error.

In the process $\bc \! \rightarrow \! \jp$, the two form factors $g$
and $f$ have been reported at non-zero recoil by one group on one
lattice volume~\cite{Colquhoun:2016osw,*ALE}.  For these form
factors, we also include an additional 20\% systematic error.  For
the other two form factors, $\mathcal{F}_1$ and $\mathcal{F}_2$, no
results have been presented.

For the final two processes $B\rightarrow D^*$ and $B_s\rightarrow
\ds^*$, only the zero-recoil value of $f$ [which is exactly related to
$\mathcal{F}_1$ at zero recoil, see above Eq.~(\ref{eq:difcofvec})] has
been computed.  In the case of $B\rightarrow D^*$, $f(t_-)$ has been
computed by two groups~\cite{Bailey:2014tva,Harrison:2017fmw}, and
here we take the FLAG value~\cite{Aoki:2019cca}.  For $B_s \!
\rightarrow \! \ds^*$, we include the recent result of
Ref.~\cite{McLean:2019sds}.  In addition to the lack of non-zero
recoil data, no results for $g$ and $\mathcal{F}_2$ at any points are
available.

The presentation of these results in the literature is also varied.
For some, only a functional form is presented; in such cases, we
resample the form factors at a fixed number of $t$ values, using the
full error estimates and correlation matrices.  Other form factors are
presented at fixed values of $t$; in such cases, we sample the form
factors at the given $t$ values.

\section{Dispersive Relations}\label{sec:da}

This work employs the model-independent form-factor parametrization
of Boyd, Grinstein, and Lebed (BGL)~\cite{Boyd:1997kz,
Grinstein:2015wqa}, which rests on the twin principles of analyticity
and unitarity of particular two-point Green's functions.   While
originally applied to the form factors of heavy-light semileptonic
decays, this parametrization was extended to heavy-heavy systems in
Refs.~\cite{Cohen:2018dgz,Berns:2018vpl} (using a slightly different
set of free parameters to simplify the computation).  The essential
ingredients are summarized here.

The two-point momentum-space Green's function $\Pi_J^{\mu \nu}$ of a
vectorlike quark current, $J^\mu \equiv \bar Q \Gamma^\mu Q^\prime
\,$, can be expanded in a variety of ways~\cite{Boyd:1994tt,
Boyd:1995cf,Boyd:1995tg,Boyd:1995sq,Boyd:1997kz}.  
For our purpose  it is
convenient is to break $\Pi_J^{\mu \nu}$ into spin-1 ($\Pi_J^T$) and
spin-0 ($\Pi_J^L$) pieces~\cite{Boyd:1997kz}.  The functions
$\Pi^{L,T}_J$ are divergent in perturbative QCD (pQCD), and require
subtractions in order to be rendered finite.  After performing the
minimum necessary numbers of subtractions for each function, one
obtains the finite dispersion relations:
\begin{eqnarray}\label{eq:chilt}
\chi^L_J (q^2) \equiv \frac{\partial \Pi^L_J}{\partial q^2} & = &
\frac{1}{\pi} \int_0^\infty \! dt \, \frac{{\rm Im} \,
  \Pi^L_J(t)}{(t-q^2)^2} \, , \nonumber \\
\chi^T_J (q^2) \equiv \frac 1 2 \frac{\partial^2 \Pi^T_J}{\partial
  (q^2)^2} & = & \frac{1}{\pi} \int_0^\infty \! dt \, \frac{{\rm Im}
  \, \Pi^T_J(t)}{(t-q^2)^3} \, .
\end{eqnarray}
Since $q^2$ remains as a free parameter in these equations, one may
select its value in order to obtain the tightest possible
phenomenological constraints (which requires that $q^2$ is as close to
the region of hadronic masses as possible), but still require that
$\chi^{\vphantom\dagger}_{J}(q^2)$ can be computed to good accuracy using pQCD (whose
asymptotic regime is the deep-Euclidean limit, $q^2 \! \to \!
-\infty$).  The parametric requirement for the latter condition is
$(m_Q + m_{Q^\prime}) \Lambda_{\rm QCD} \ll (m_Q + m_{Q^\prime})^2 \!
- \! q^2$, which is clearly satisfied by $q^2 \! = \! 0$ for any
process in which either or both of $Q$, $Q^\prime$ is heavy compared
to $\Lambda_{\rm QCD}$, as is true for all cases considered here.
$\chi^{\vphantom\dagger}_J (q^2=0)$ has been computed to two-loop
pQCD order, including leading nonperturbative vacuum
condensates~\cite{Generalis:1990id,Reinders:1980wk,Reinders:1981sy,
Reinders:1984sr,Djouadi:1993ss,Bigi:2016mdz}.

Unitarity requires that each of the functions ${\rm Im} \, \Pi_J$
admits an expansion over all hadronic states $X$ that couple to the
vacuum through the current $J^\mu$:
\begin{equation} \label{eq:FullPi}
{\rm Im} \, \Pi^{T,L}_J (q^2) = \frac 1 2 \sum_X (2\pi)^4 \delta^4
(q - p_X) \left| \left< 0 \! \left| J \right| \! X \right> \right|^2
\, .
\end{equation}
Since every nontrivial term in Eq.~(\ref{eq:FullPi}) is positive, one
may truncate the sum on X after any number of states, insert the sum 
into Eqs.~(\ref{eq:chilt}), and obtain a strict inequality based upon
unitarity.  While typically these inequalities have been employed for
single states $X$, they clearly become stronger when more states $X$
are included~\cite{Boyd:1997kz,Bigi:2017jbd,Jaiswal:2017rve}, which
is a key ingredient of our analysis here.  Our set of $X$ includes
only below-threshold $B_c^{(*)}$ poles and the two-body channels
discussed above.  Additional branch points corresponding to the
thresholds of processes such as $B_c \pi \pi$ occur at lower $t$
values, but their contributions to the dispersive bounds are expected
to be small due to OZI suppression, closeness to the $B^{(*)}D$
thresholds that are already taken into account, or both.

For the purposes of this work, the first physically significant
two-body production threshold occurs at $t \! = \! t_{\rm bd} \!
\equiv \!  (M_{B^{(*)}} \! \!  + \! M_{D})^2$, depending upon which
component of the two-point function is being considered (see
Table~\ref{tab:poles}).  $t_{\rm bd}$ thereby represents the lowest
significant branch point in the two-point function.

Analyticity properties of the Green's function are incorporated by
a conformal mapping of the complex-$t$ plane with a cut beginning at
the branch point $t \! = \! t_*$ to the unit disk in a complex
variable $z$:
\begin{equation} \label{eq:zdef}
z(t;t_0) \equiv \frac{\sqrt{t_* - t} - \sqrt{t_* - t_0}}
{\sqrt{t_* - t} + \sqrt{t_* - t_0}} \, ;
\end{equation}
the two edges of the branch cut in $t$  are mapped to the
unit circle $C$ in $z$.
The parameter $t_0$ is free at this stage; we later optimize this
choice [Eqs.~(\ref{eq:Ndefn})--(\ref{eq:zmaxminopt})] to improve the
convergence of the Green's function in the variable $z$.

The importance of allowing a branch point $t_*$ that does not
necessarily equal $t_+$ becomes apparent in processes for which $t_+$
lies well above the lowest significant branch point for the
two-point function.  Such an effect is especially significant for
baryon decays such as $\Lambda_b\rightarrow\Lambda_c$ and
$\Lambda_b\rightarrow p$.  Previous studies that automatically set
$t_* \! = \! t_+$~\cite{Boyd:1995tg,Boyd:1997kz} can introduce branch
cuts ({\it e.g.}, for $B^{(*)}D^{(*)}$ pairs) inside the unit circle
$C$ defined by $|z| \! = \! 1$.  The purpose of the BGL
parametrization being to eliminate all significant 
nonanalytic behavior below-$t_*$, there are two choices: either model
the strength of the branch cut (which requires both knowledge of the
branch point and the function along the cut) and use this information
to loosen the strength of the bounds, or instead  set
$t_* \! = \! t_{\rm bd}$ (as is done here).  With this latter choice,
$t_* \! < \! t_+$ is no longer the threshold relevant to the physical
process, but it is the location of an important branch cut in the
two-point function to which the process contributes.  Nevertheless,
for all heavy-quark systems, one finds that the semileptonic decay
region $m_\ell^2\leq t\leq t_-$ lies substantially below
$t_{\rm bd}$, and therefore the BGL bounds are not strongly affected.

With this choice, the bounds obtained by inserting
Eq.~(\ref{eq:FullPi}) into Eqs.~(\ref{eq:chilt}) amount to an integral
over the unit circle $|z| \! = \! 1$ of an integrand containing the
form factor $F_i$ of interest multiplied by the known {\it outer
functions\/} $\phi_i(t;t_0)$, which incorporate information about
kinematics and changes-of-variable.   (These functions are tabulated
for the cases of interest in
Refs.~\cite{Cohen:2018dgz,Berns:2018vpl}.)  The only significant
nonanalytic features remaining within the unit circle $|z| \! = \! 1$
are simple poles corresponding to $B_c^{(*)}$ states.  Each such a
pole at a known location $t \! = \! t_s$ can effectively be removed
from the integrand through multiplication by $z(t;t_s)$ (a {\it
Blaschke factor}).  In the $b \! \to \! c$ processes of interest here, 
the masses corresponding to the $B_c^{(*)}$ poles that
must be removed in this analysis are collected in
Table~\ref{tab:poles}, organized by the $J^P$ channel to which each
one contributes ($1^- \, \{f_+, \, g, F_V,H_V\}$; $1^+ \, \{f, {\cal
F}_1, G_1, H_A\}$; $0^+ \, \{f_0, F_0\}$; $0^- \, \{{\cal F}_2,
G_0\}$).  These masses have either been measured at the
LHC~\cite{Aaij:2016qlz,Sirunyan:2019osb} (boldface) or are derived
from very recent model calculations~\cite{Eichten:2019gig}.

\begin{table}
\caption{\label{tab:poles}All $B_c$-state masses lying below the
thresholds $t \! = \! t_{\rm bc}$ (for which ``Lowest pair''
indicates the states whose masses enter into $t_{\rm bc}$) for the
$J^P$ channels relevant to this study.  Numbers in bold are masses
measured at the LHC.}
\begin{center}
\begin{tabular}
{l c c c}
\hline\hline
Type &  $J^P$&Lowest pair  & $M$ [GeV]\\
\hline
Vector & $1^-$&$BD$&6.3290, 6.8975, 7.0065\\
\hline
Axial & $1^+$ &$B^*D$&6.7305, 6.7385, 7.1355, 7.1435\\
\hline
Scalar&$0^+$&$BD$&6.6925, 7.1045\\
\hline
Pseudoscalar&$0^-$&$B^*D$&\textbf{6.2749(8)}, \textbf{6.8710(16)}\\
\hline
\end{tabular}
\end{center}
\end{table}

Denoting the product of Blaschke factors for all poles with $|z| \!
< \! 1$ as $P_i$, the unitarity bound for the form factor $F_i$
expressed entirely in terms of the conformal variable $z$ reads
\begin{equation} \label{eq:FFrelnz}
\frac{1}{2\pi i} \sum_i\oint_C \frac{dz}{z}
| \phi_i(z) P_i(z) F_i(z) |^2 \le 1 \, .
\end{equation}
Since the product $\phi_i(z) P_i(z) F_i(z)$ is an analytic function
inside the unit circle $|z| \! = \! 1$, one may write
\begin{equation} \label{eq:param}
F_i(t) = \frac{1}{|P_i(t)| \phi_i(t;t_0)} \sum_{n=0}^\infty a^i_{n}
z(t;t_0)^n\, .
\end{equation}
Inserting Eq.~(\ref{eq:param}) into Eq.~(\ref{eq:FFrelnz}), one finds
that the unitarity bound can be compactly written as a constraint on
the Taylor-series coefficients $a^i_n$:
\begin{equation} \label{eq:coeffs}
\sum_{i;n=0}^\infty (a^i_{n})^2 \leq 1 \, .
\end{equation}
Equations~(\ref{eq:param}) and (\ref{eq:coeffs}) are the essence of
the BGL pa\-rametrization.  Every functional form for $F_i(t)$ that
respects analyticity and unitarity, as expressed by
Eqs.~(\ref{eq:chilt}) and (\ref{eq:FullPi}), can be expressed in terms
of a set of Taylor coefficients $a^i_{n}$ that satisfy the sum rule
Eq.~(\ref{eq:coeffs}).

As in Ref.~\cite{Cohen:2018dgz}, the generalization of the location
of the branch point from $t_+$ to $t_{\rm bd}$ means that slightly
more complicated functions of the mass parameters appear in the
analysis.  Reprising this previous notation, we define
\begin{align}
 r \equiv &\frac{m}{M} , \phantom{xxx}\delta \equiv \frac{m_\ell}{M}
 , \nonumber\\
 \beta \equiv &\frac{M_{B^{(*)}}}{M} , \phantom{xx}\Delta \equiv
\frac{M_{D}}{M} , \nonumber\\
 \kappa \equiv &(\beta+\Delta)^2-(1-r)^2 , \nonumber\\
 \lambda \equiv &(\beta+\Delta)^2-\delta^2 , \label{eq:newparams}
\end{align}
and the free parameter $t_0$ in Eq.~(\ref{eq:zdef}) is replaced by a
parameter $N$:
\begin{equation} \label{eq:Ndefn}
N \equiv \frac{t_{\rm bd} - t_0}{t_{\rm bd} - t_-} \, .
\end{equation}
Computing the kinematical range for the semileptonic process given in
terms of $z$ is then straightforward.  The minimal (optimized)
truncation error is achieved when $z_{\rm min} = -z_{\rm max}$, which
occurs when $N \! = \! N_{\rm opt} = \sqrt{\frac{\lambda}{\kappa}}$.
At this point, one obtains
\begin{equation} \label{eq:zmaxminopt}
  z_{\rm max} = -z_{\rm min} = \frac{\lambda^{1/4} \! -\kappa^{1/4}}
  {\lambda^{1/4} \! +\kappa^{1/4}}\, .
\end{equation}
One finds that the semileptonic decay processes under consideration
do not exceed $z_{\rm max}\approx0.03$.  If instead $t_{\rm bd}$ is
set equal to $t_+$, then Eqs.~(\ref{eq:newparams}) reduce to
$\Delta\rightarrow r$, $\beta\rightarrow 1$, and $\kappa \! \to \!
4r$, and all of the expressions reduce to those given in
Ref.~\cite{Grinstein:2015wqa}.

\section{Results}\label{sec:results}
The global analysis of the $b\rightarrow c$ hadronic form factors
relies upon a number of constraints.  They are summarized here for
the convenience of the reader:
\begin{itemize}
 \item The $n\leq2$ coefficients $a_n^{i}$ in each channel are
 constrained by $\sum_{n,i}(a_n^{i})^2\leq 1$
 [Eq.~(\ref{eq:coeffs})]. 
 \item The form factors satisfy the exact kinematic relations below
 Eq.~(\ref{eq:f0}), above Eq.~(\ref{eq:difcofvec}), and below
 Eqs.~(\ref{eq:BaryFF})]:
 \begin{align}
\mathcal{F}_1(t_-)&=M(1-r)f(t_-) \, , \nonumber\\ 
\mathcal{F}_1(0) \! &= \! \frac 1 2 M^2(1-r^2)\mathcal{F}_2(0) \, ,
\nonumber\\
f_+(0)&=M^2(1-r^2)f_0(0) \, , \nonumber\\
H_A(t_-)&=M(1-r)G_1(t_-) \, , \nonumber\\
M(1+r)F_0(0)&=M(1-r)H_V(0) \, \nonumber\\
M(1-r)G_0(0)&=M(1+r)H_A(0) \, .
\end{align}
 \item $B_{(s)} \! \rightarrow \! D_{(s)}^*$ form factors are taken 
to be consistent with the form factor $\xi_{(s)}(w)$ derived from
 $B_{(s)} \! \rightarrow \! D_{(s)}$,   once an
 additional 20\% systematic error is included 
 to account for violations of Isgur-Wise scaling.
 \item $B_{(s)} \! \rightarrow \! D_{(s)}^*$ form factors are maximal
 at the zero-recoil point $t \! = \! t_-$, since the universal form
 factor $\xi_{(s)}$ represents an overlap matrix element between
 initial and final states. This condition is implemented via the
 constraints $F_i(t_-) \! \geq \!  F_i(0)$ and $\frac{d F_i}{d t}
 \big|_{t_-} \! \geq \! 0$, where $F_i$ represents any of the form
 factors.
 \item The normalizations and slopes of the $\bc\rightarrow \! \{\jp,
 \ec\}$ form factors $f_+,g \, (\propto \! V),\mathcal{F}_1,
 \mathcal{F}_2$ are required to be consistent at zero recoil [via
 Eqs.~(\ref{eq:hhdelta})] to  within 50\% with the results for $f_0,f \,
 (\propto \! A_1)$ computed from lattice QCD\@.
 
\end{itemize}

\begin{figure}
 \includegraphics[width=\linewidth]{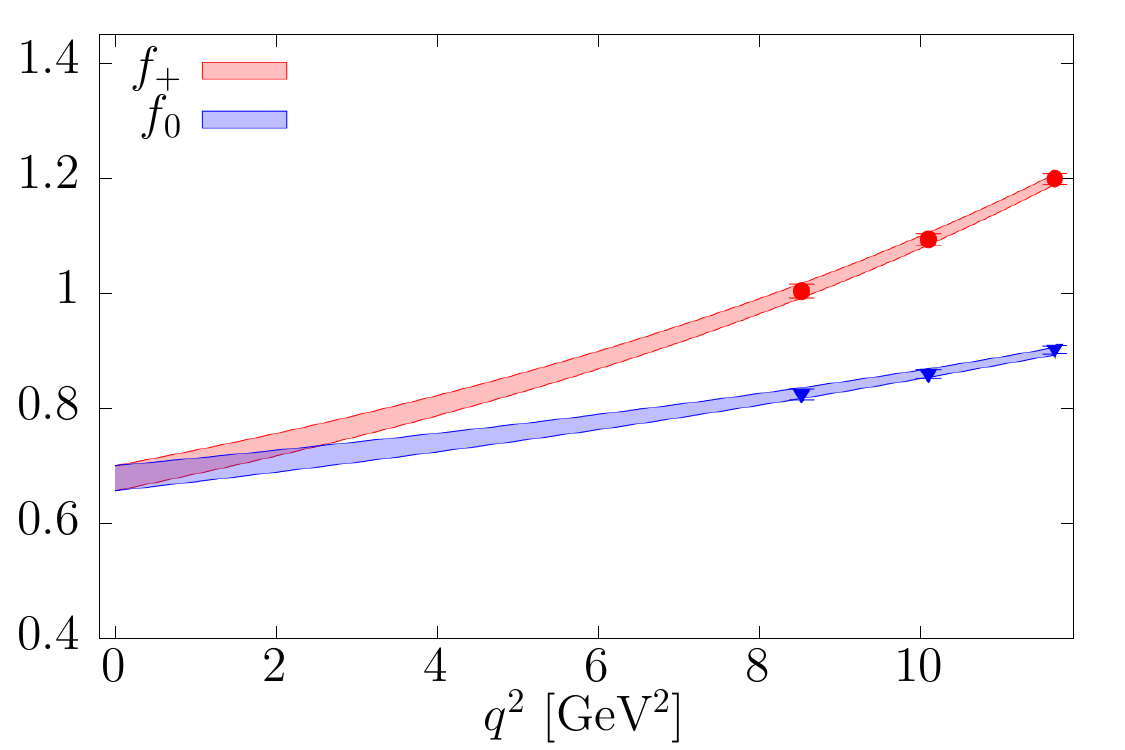}
 \caption{\label{fig:latff1}$B\rightarrow D$ form factors: $f_+(q^2)$
(red circles) and $f_0(q^2)/M^2(1-r^2)$ (blue triangles)
from~\cite{Lattice:2015rga}.  The colored bands are the
one-standard-deviation ($1\sigma$) best-fit regions obtained from our
global dispersive analysis.}
\end{figure}

\begin{figure}
 \includegraphics[width=\linewidth]{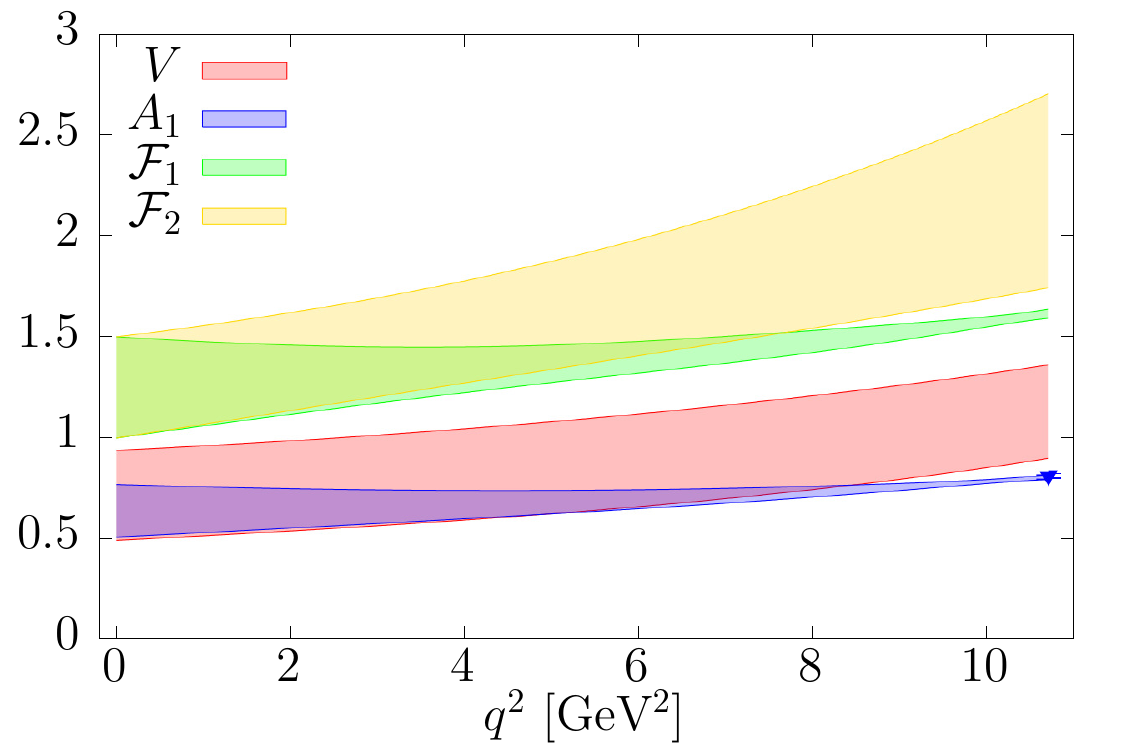}
\caption{\label{fig:latff2}$B\rightarrow D^*$ form factors:
$A_1(q_{\rm max}^2)$ (blue triangle) from \cite{Aoki:2019cca}.  The
colored bands are the one-standard-deviation ($1\sigma$) best-fit
regions obtained from our global dispersive analysis. $\mathcal{F}_1$
has been divided by $\frac{1}{2}M^2(1-r^2)$.}
\end{figure}

\begin{figure}
 \includegraphics[width=\linewidth]{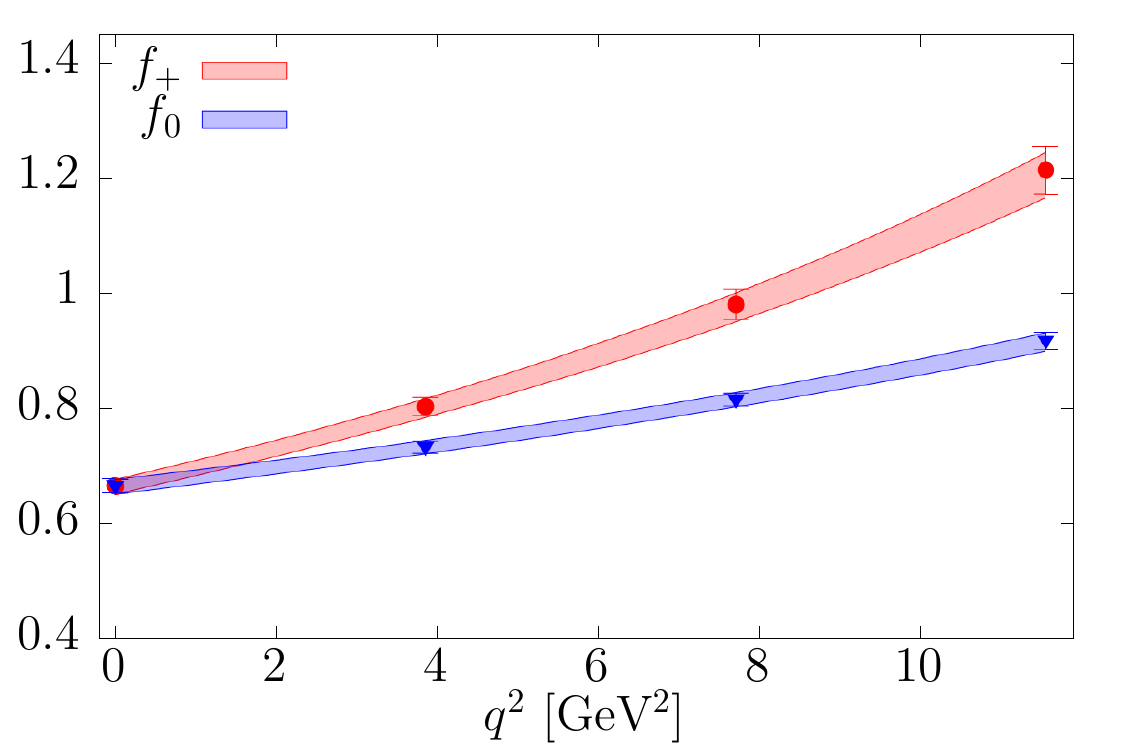}
 \caption{\label{fig:latff3}$B_s\rightarrow D_s$ form factors:
$f_+(q^2)$ (red circles) and $f_0(q^2)/M^2(1-r^2)$ (blue triangles)
from~\cite{McLean:2019qcx}.  The colored bands are the
one-standard-deviation ($1\sigma$) best-fit regions obtained from our
global dispersive analysis.}
\end{figure}

\begin{figure}
 \includegraphics[width=\linewidth]{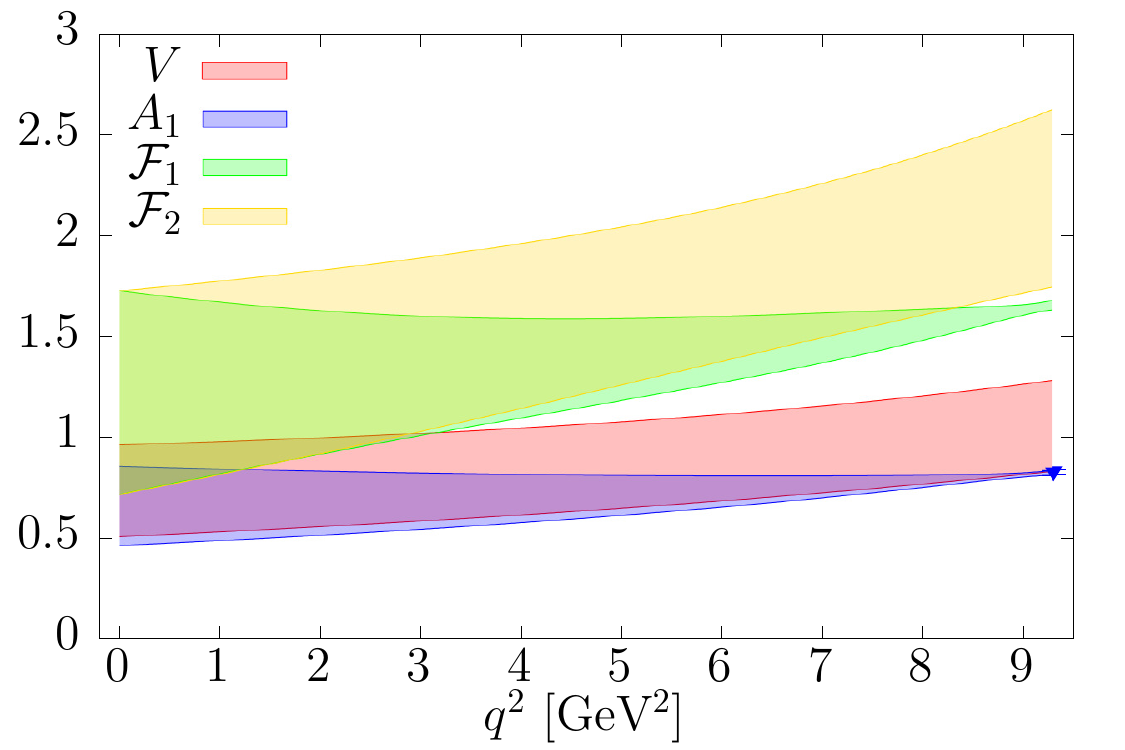}
 \caption{\label{fig:latff4}$B_s\rightarrow D_s^*$ form factors:
 $A_1(q_{\rm max}^2)$ (blue triangle) from \cite{McLean:2019sds}.
 The colored bands are the one-standard-deviation ($1\sigma$)
 best-fit regions obtained from our global dispersive analysis.
 $\mathcal{F}_1$ has been divided by $\frac{1}{2}M^2(1-r^2)$.}
\end{figure}

\begin{figure}
 \includegraphics[width=\linewidth]{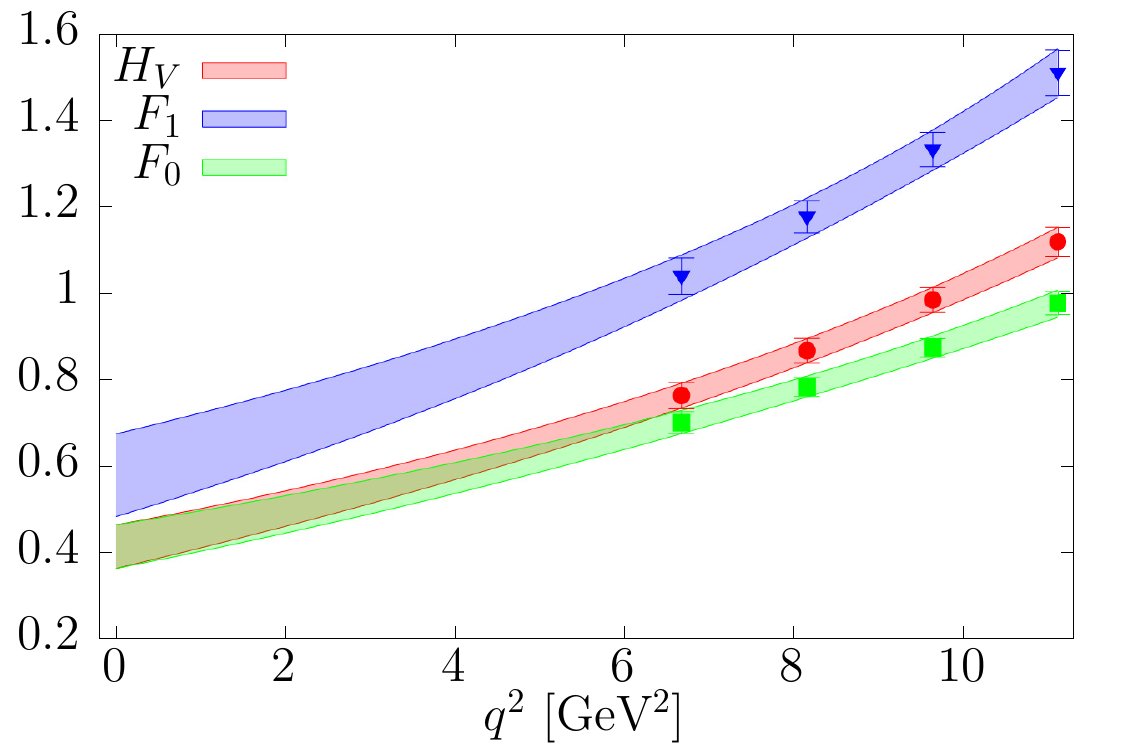}
 \caption{\label{fig:latff5}$\Lambda_b\rightarrow \Lambda_c$ form
factors: $H_V(q^2)/M(1+r)$ (red circles), $F_1(q^2)$ (blue
triangles), and $F_0(q^2)/M(1-r)$ (green squares)
from~\cite{Detmold:2015aaa}.  The colored bands are the
one-standard-deviation ($1\sigma$) best-fit regions obtained from our
global dispersive analysis.}
\end{figure}

\begin{figure}
 \includegraphics[width=\linewidth]{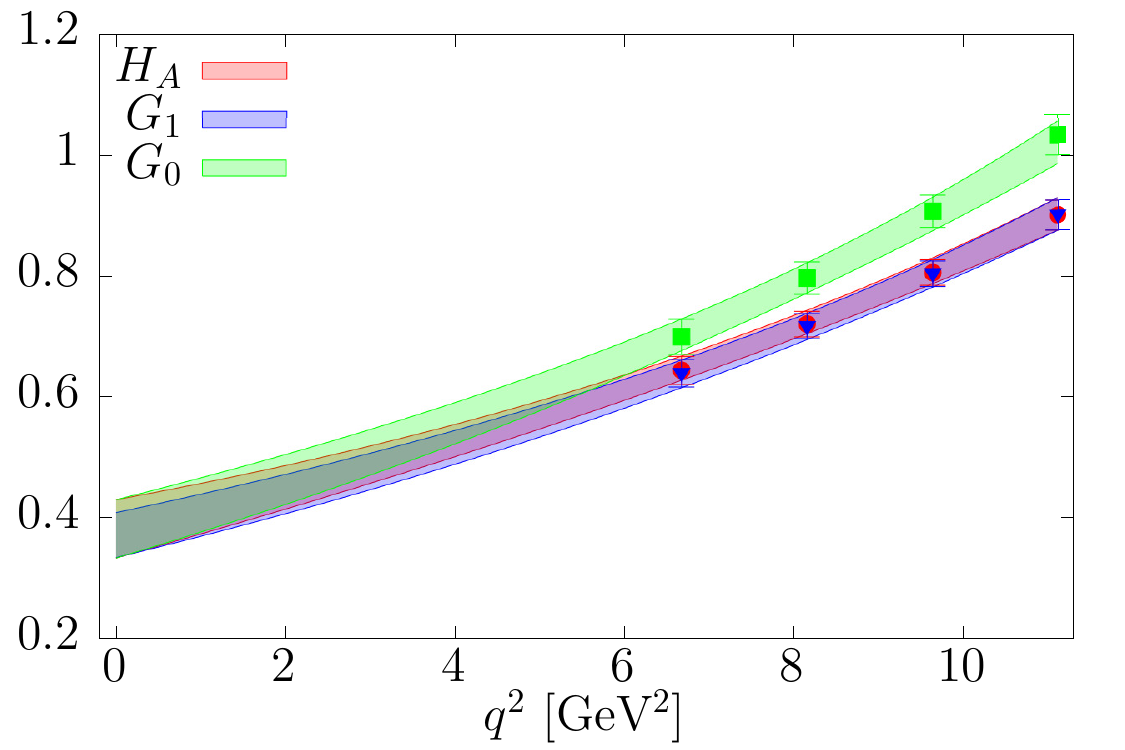}
 \caption{\label{fig:latff6}$\Lambda_b\rightarrow \Lambda_c$ form
factors: $H_A(q^2)/M(1-r)$ (red circles), $G_1(q^2)$ (blue
triangles), and $G_0(q^2)/M(1+r)$ (green squares)
from~\cite{Detmold:2015aaa}.  The colored bands are the
one-standard-deviation ($1\sigma$) best-fit regions obtained from our
global dispersive analysis.}
 \end{figure}

\begin{figure}
 \includegraphics[width=\linewidth]{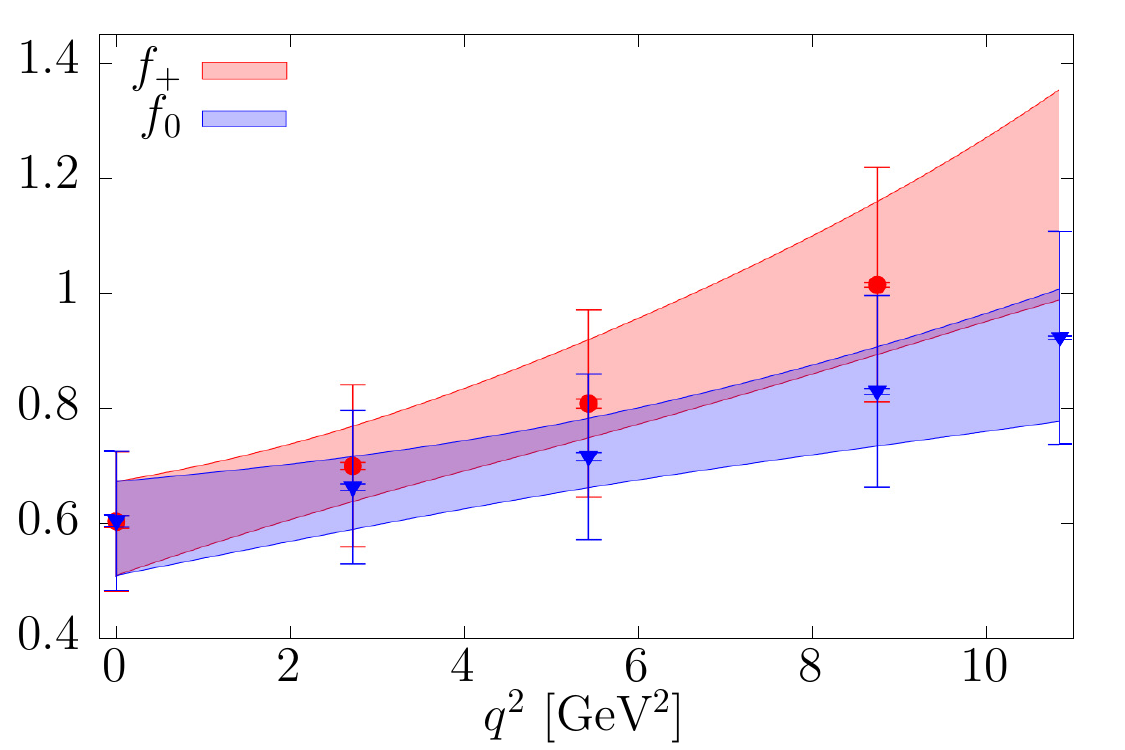}
 \caption{\label{fig:latff7}$\bc\rightarrow \ec$ form factors:
$f_+(q^2)$ (red circles) and $f_0(q^2)/M^2(1-r^2)$ (blue triangles)
from the HPQCD Collaboration~\cite{Colquhoun:2016osw,ALE}.  The
interior bars represent the statistical uncertainty quoted by
HPQCD\@.  The exterior bars represent the result of including our
$f_{\rm lat} \! = \! 20\%$ systematic uncertainty.  The colored bands
are the one-standard-deviation ($1\sigma$) best-fit regions obtained
from our global dispersive analysis.}
\end{figure}

\begin{figure}
 \includegraphics[width=\linewidth]{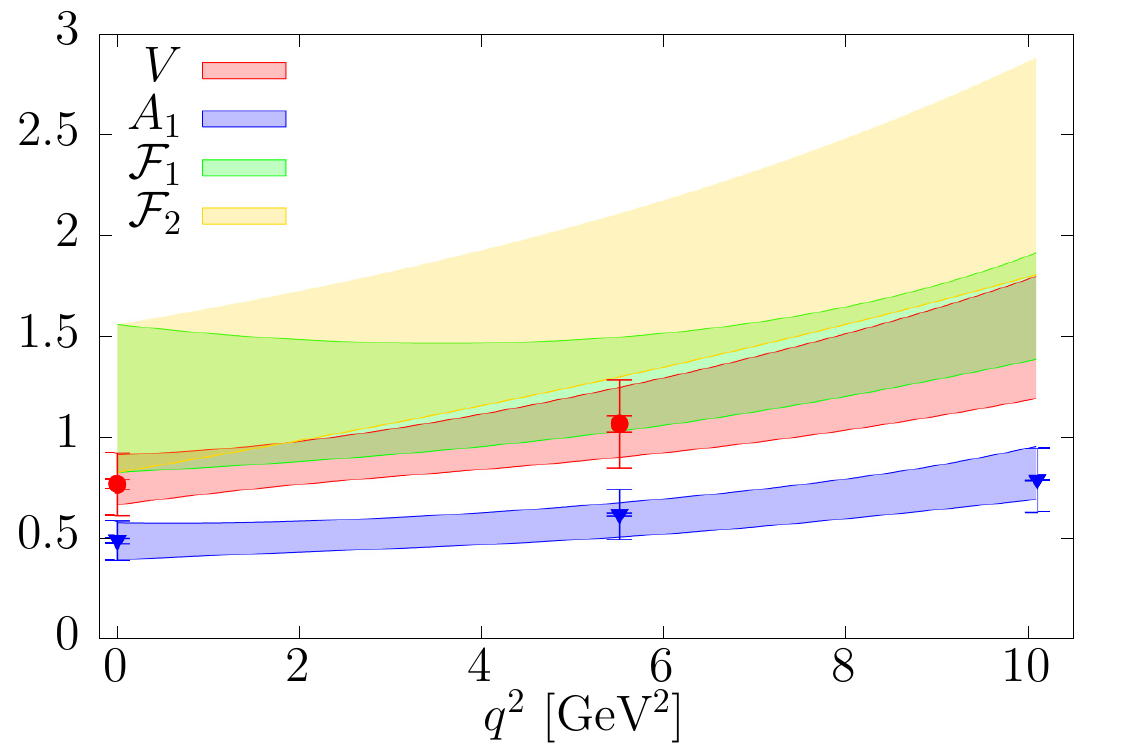}
 \caption{\label{fig:latff8}$\bc\rightarrow \jp$ form factors:
$V(q^2)$ (red circles) and $A_1(q^2)$ (blue triangles) from the HPQCD
Collaboration~\cite{Colquhoun:2016osw,ALE}.  The interior bars
represent the statistical uncertainty quoted by HPQCD\@.  The
exterior bars represent the result of including our
$f_{\rm lat} \! = \! 20\%$ systematic uncertainty.  The colored bands
are the one-standard-deviation ($1\sigma$) best-fit regions obtained
from our global dispersive analysis.  $\mathcal{F}_1$ has been
divided by $\frac{1}{2}M^2(1-r^2)$.}
\end{figure}

We perform the constrained multivariable fit by first randomly
sampling $q^2$ values of the form factors for which lattice data is
known.  If correlations have been reported by the lattice QCD groups,
they are implemented in the sampling.  The preliminary HPQCD results
for $\bc\rightarrow\{\jp,\ec\}$ report only statistical error.  To
account for the unknown systematics like finite-volume and
discretization effects, we include in quadrature an additional systematic
error $f_{lat}=20\%$ (as a percentage of the form factor at each
point).  Lines of best fit are then computed from the collection of
sampled points using a least-squares procedure.  The resulting form
factors, exhibited with one-$\sigma$ bands, are presented in
Figs.~\ref{fig:latff1}--\ref{fig:latff8}.  The $a_n^i$ for the form
factors can be found in Table~\ref{tab:coef}.  Of particular note,
our theoretical values for the form factors are consistent with those
of the two processes $B\rightarrow D^{(*)}$ for which experimental
data has been obtained~\cite{Aubert:2009ac,Glattauer:2015teq,
Dey:2019bgc}.

In interpreting these $1\sigma$ bands, it is important to recall
that this analysis includes statistical errors from the lattice
studies  for which the notion of ``$1\sigma$''  is well defined.
However, the analysis also includes systematic errors for which,
strictly speaking, it is not.  Moreover, in assigning systematic
error associated with limited lattice data for which systematic
errors had not been carefully studied, or due to truncation errors in
the theory, we have been quite conservative.  Thus, one might
reasonably expect the SM result to fall within these bands with a
higher probability than had the bands been entirely due to
statistical errors.

From Table~\ref{tab:coef}, it is possible to investigate the
convergence of BGL expansion.  All the $a_2$ coefficients are
consistent with zero at $1.2\sigma$, suggesting that each series is
rapidly converging; additional parameters are unnecessary at the
present precision of lattice data.  This lack of precision also
allows for the $a_2$ parameters to fluctuate substantially, such that
in a given fit each one can typically be $\mathcal{O}(0.1)$, and
therefore contribute significantly to the dispersive bounds of
Eq.~(\ref{eq:coeffs}).

With this observation, one would expect the dispersive bounds to be
saturated, similar to the results of~\cite{Cohen:2018dgz} in which
the dispersive bound for the unknown form factor $\mathcal{F}_2$  was
saturated.  Fitting all seven processes together,
Eq.~(\ref{eq:coeffs}) is typically saturated for all four channels
($T, L$; $V, A$).  However, this result occurs not only because the
$a_2$ parameters are not well constrained.  Surprisingly,
Table~\ref{tab:coef} shows that for the two $\lb\rightarrow\lc$ form
factors $F_0,G_0$, the $a_1$ coefficients are $\mathcal{O}(0.5)$.
Each one of them saturates about 25\% of the dispersive bound.  This
result suggests that the inclusion of baryonic channels into the
dispersive approach is particularly powerful.

 In the case of $G_0$, there are additional benefits beyond providing 
such a large contribution toward saturation. In the $0^-$ channel,
only the $G_0$ and $\mathcal{F}_2$ form factors contribute.  At
present, no lattice results for any $\mathcal{F}_2$ exist.  Given
that $\mathcal{F}_2$ form factors contribute significantly only to
$\tau$ decays, this uncertainty is a sizeable fraction of the
uncertainty in our predictions of $R(H_c)$. The large contribution of
$G_0$ to the dispersive bound reduces this error.  These dual
benefits from including $\lb\rightarrow\lc$ should motivate future
efforts to obtain lattice results for form factors of other baryonic
processes, {\it i.e.}, $\lb\rightarrow\lc^*$.

\begin{table}
\caption{\label{tab:coef}The BGL coefficients $a_n^{i}$ of this $N=2$
analysis.  The full correlation matrix between the coefficients can
be found in the supplemental material.}
\begin{center}
\begin{tabular}
{l l| c c c}
\hline\hline
$b\rightarrow c$ & $F_i$ & $a_0^{i}$ $[\times10^2]$ & $a_1^{i}$
$[\times10^2]$& $a_2^{i}$ $[\times10^2]$\\
\hline
$B\rightarrow D$ &$f_0$ & 7.2(1.0) & --17.(4) & 0.7(6)\\ 
&$f_+$ & 0.25(4) & --0.58(9) & 1(14)\\
\hline
$B\rightarrow D^*$  &$g$ & 0.67(17) & --0.5(14) & 10(40)\\ 
&$f$ & 0.42(3) & --0.2(13) & 20(30)\\ 
&$\mathcal{F}_1$ & 0.07(4) & --0.11(17) & 1(7)\\ 
&$\mathcal{F}_2$ & 5.4(1.0) & --20(30) & --10(60)\\ 
\hline
$B_s\rightarrow D_s$&$f_0$ & 5.23(8) & --16(17) & 1(4)\\ 
&$f_+$ & 0.179(5) & --0.47(9) & --1.6(5)\\ 
\hline
$B_s\rightarrow D_s^*$  &$g$ & 0.46(11) & 0(300) & 16(16)\\ 
&$f$ & 0.33(4) & --0.4(20) & 18(19)\\ 
&$\mathcal{F}_1$ & 0.05(6) & --0.2(4) & 2(11)\\ 
&$\mathcal{F}_2$ & 4(9) & --20(30) & 0(300)\\ 
\hline
$\Lambda_b\rightarrow \Lambda_c$&$H_V$ & 0.256(9) & --2.7(4) & 2(7)\\ 
&$F_1$ & 0.85(4) & --7.9(1.7) & 4(14)\\ 
&$F_0$ & 5.14(18) & --46(9) & 1(3)\\ 
&$H_A$ & 0.0613(18) & --0.49(9) & --1(3)\\ 
&$G_1$ & 0.356(12) & --2.7(5) & --3(4)\\ 
&$G_0$ & 5.23(18) & --53(9) & 1.3(1.2)\\ 
\hline
$\bc\rightarrow\ec$ &$f_0$ & 6.1(6) & --30(30) & 10(70)\\ 
&$f_+$ & 0.18(3) & --0.8(7) & --3(15)\\ 
\hline
$\bc\rightarrow\jp$ &$g$ & 0.47(9) & --2(3) & 20(60)\\ 
&$f$ & 0.34(5) & --2.6(2.0) & 40(60)\\ 
&$\mathcal{F}_1$ & 0.058(10) & --0.3(3) & 9(8)\\ 
&$\mathcal{F}_2$ & 4(10) & --21(16) & 0.8(9)\\ 

\hline
\end{tabular}
\end{center}
\end{table}

In the case of $\lb\rightarrow\lc$, sufficient lattice data exists so
that the constraint of heavy-quark symmetries  is not required to fix
the form factors.  Therefore, we can use our results in that process
to investigate how well the HQET relations are satisfied.  While
higher-order terms are known~\cite{Bernlochner:2018kxh}, we consider
the leading-order relations in which the six form factors are all
proportional to an Isgur-Wise function $\zeta(w)$, which is typically
expanded in powers of $w-1$ as
\begin{equation}
 \zeta(w)=\zeta(1)-\rho^2(w-1)+\frac{1}{2}\sigma^2(w-1)^2.
\end{equation}
In this expansion, our results for the coefficients of the Taylor
series are found in Table~\ref{tab:lblc}.  One can see that our
results, despite suggesting corrections to the HQET relations, are
consistent with the sum-rule bounds: $\rho^2\geq0$ and $\sigma^2\geq
\frac{3}{5}\left[\rho^2+(\rho^2)^2\right]$~\cite{LeYaouanc:2008pq}.

In the final two rows of Table~\ref{tab:lblc}, we compute a pair of
averages of the coefficients.  The first, $\zeta_{\rm AVG}$, is simply
an average of parameters from all six form factors together, and would
represent a best-fit phenomenological value for $\zeta(w)$.  The last
row ($\zeta_{w \to 1}$) instead averages over only $H_A$ and $G_1$.
This average is of interest because only these two form factors
contribute appreciably to the differential decay rate of
$\lb\rightarrow\lc\mu^-\bar{\nu}_{\mu}$, the process measured by the
LHCb Collaboration~\cite{Aaij:2017svr}.  In that work, assuming the
same leading-order HQET relations and the static approximation, LHCb
extracted the values $\rho^2=1.63(11)$ and $\sigma^2=2.16(34)$ from
the decay $\lb\rightarrow\lc\mu^-\bar{\nu}_{\mu}$.  Good agreement is
found between the LHCb results and those of $\zeta_{\rm AVG}$.  But if
these results were used to make predictions far from $w\rightarrow 1$,
or for $\lb\rightarrow\lc\tau^-\bar{\nu}_{\tau}$, then the other form
factors begin to contribute appreciably, and a systematic error would
be introduced because their corresponding coefficients differ
dramatically from those of $\zeta_{w \to 1}$.

\begin{table}
\caption{\label{tab:lblc}HQET expansion parameters for
$\lb\rightarrow \lc$ obtained from this analysis.  $\zeta_{\rm AVG}$
indicates the values of the Isgur-Wise function obtained by averaging
all of the form factors, while $\zeta_{w\rightarrow 1}$ is obtained
by averaging only over the form factors $H_A,$ $G_1$ that contribute
significantly at zero recoil.}
\begin{center}
\begin{tabular}
{l | c c c}
\hline\hline
$F_i(q^2)$ & $F_i(1)$ & $\rho^2$ & $\sigma^2$\\
\hline
$H_V$&1.12(4)&2.5(3)&5.6(1.8)\\
$F_1$&1.51(7)&3.3(5)&8.0(1.8)\\
$F_0$&0.97(4)&1.8(3)&3.6(6)\\
$H_A$&0.90(3)&1.7(3)&3.4(1.8)\\
$G_1$&0.90(3)&1.82(18)&4.0(8)\\
$G_0$&1.02(4)&2.2(3) & 4.8(6)\\
\hline
$\zeta_{\rm AVG}$&1.1(3)&2.2(7)&4(2)\\
\hline
$\zeta_{w\rightarrow1}$&0.90(3)&1.7(2)&3.6(1.4)\\
\hline
\end{tabular}
\end{center}
\end{table}

Using the computed form factors, we extract three observables of
experimental interest, and present the results in
Table~\ref{tab:results}.  The first is the semileptonic decay ratio:
\begin{equation}
 R(H_c)=\frac{\Gamma(H_b\rightarrow H_c \tau {\nu}_\tau)}
{\Gamma(H_b\rightarrow H_c \mu {\nu}_\mu)}.
\end{equation}
For those $R(H_c)$ for which existing theoretical values exist, we
find good agreement.  This result is to be expected, given that all
the theoretical values rely at least in part upon the same
lattice-QCD data used here.  Beyond these checks, we have produced
two new SM predictions, those of $\rjp=\rjpv$ and
$R(\ds^*)=\rdsstv$, which can be compared to the upcoming LHCb
results of Runs II and III\@. We find that the $\rjp$ prediction is within $1.8\sigma$ of 
the current LHCb result of
$0.71(17)(18)$~\cite{Aaij:2017tyk}.

The second observable is the polarization of the $\tau$ lepton, given
by
\begin{equation}
 P_\tau(H_c) \equiv \frac{\Gamma^+-\Gamma^-}{\Gamma^++\Gamma^-} \, ,
\end{equation}
where $\Gamma^{\pm}$ are the decay rates of a $\tau$ with fixed
helicity $\lambda=\pm$.  Only $P_\tau(D^*)=-0.38(60)$ has been
measured~\cite{Hirose:2016wfn,Hirose:2017dxl} and our value
$-0.51(15)$ agrees within uncertainties.  For the other processes, we
present predictions for comparison with future measurements. 

The final observable we compute is the fractional longitudinal
polarization of the decaying vector meson:
\begin{equation}
 F_L^{H_c} \equiv \frac{\Gamma^0}{\Gamma} \, ,
\end{equation}
where $\Gamma^{0}$ is the decay rate of a vector $H_c$ with helicity
$\lambda \! = \! 0$.  In the case of the $D^*$, this quantity has
been measured to be $F_L^{D^*}=0.60(9)$~\cite{Abdesselam:2019wbt},
which is within $1.6\sigma$ of our result and other existing SM
values~\cite{Huang:2018nnq,Bhattacharya:2018kig}.

\begin{table}
\caption{\label{tab:results}Results of our dispersive analysis for
the semileptonic decay ratio $R(H_c)$, $\tau$ polarization
$P_\tau(H_c)$, and the (vector) $H_c$ polarization fraction
$F_L^{H_c}$.}
\begin{center}
\begin{tabular}
{l| c c c}
\hline\hline
$b\rightarrow c$ & $R(H_c)$ & $P_\tau(H_c)$& $F_L^{H_c}$\\
\hline
$B\rightarrow D$&\rdv&\pdv&---\\
$B\rightarrow D^*$&\rdstv&\pdstv&\fdstv\\
$\bs\rightarrow \ds$&\rdsv&\pdsv&---\\
$\bs\rightarrow \ds^*$&\rdsstv&\pdsstv&\fdsstv\\
$\lb\rightarrow\lc$&\rlcv&\plcv&---\\
$\bc\rightarrow\ec$&\recv&\pecv&---\\
$\bc\rightarrow\jp$&\rjpv&\pjpv&\fjpv\\
\hline
\end{tabular}
\end{center}
\end{table}

\section{Discussion and Conclusion}\label{sec:con}

In this work we have presented model-independent predictions of the
$b\rightarrow c$ hadronic transition form factors for the processes,
$B \! \to \! \{D,D^*\}$, $B_s \! \to \! \{D_s,D_s^{*}\}$, $B_c \! \to
\! \{ \jp, \ec \}$, and $\Lambda_b \! \to \! \Lambda_c$, using a
coupled global analysis.  From these form factors, Standard-Model
values for $R(H_c)$ ($\tau$-$\mu$ ratio), $P_\tau(H_c)$ ($\tau$
polarization), and $F^{H_c}_L$ ($H_c$ longitudinal component) were
computed.  Also obtained, for the first time using this approach, are
$R(D_s^*)=\rdsstv$ and $\rjp=\rjpv$.  The near-term outlook for
higher-statistics measurements from BELLE and LHCb, coupled with new
lattice results, promise to reduce the uncertainty on the
experimental and theoretical values dramatically, allowing for a
refinement of the investigation of the charged-current anomalies.
Additionally, new measurements like $R(D_s^{(*)})$ can be compared to
our results to provide complementary constraints.

We have also derived nonzero recoil relations between the heavy-heavy
meson form factors $B_c \! \to \! \{ \jp, \ec \}$ and the
Isgur-Wise-like form factor $\Delta$ at leading order in NRQCD\@.
These results allow constraints on the slopes of unknown lattice form
factors at $w=1$ to be obtained.  Furthermore, these relations can be
used as the basis of phenomenological models for the form factors.

The dominant sources of uncertainty in this analysis arise from the
form factors for which no lattice data has been reported, all of
which are in the $P\rightarrow V$ processes.  Upcoming results for
$B\rightarrow D^*$~\cite{Vaquero:2019ary} and $\bc\rightarrow\jp$
promise to provide insight into these form factors.  The global
analysis could also benefit from the inclusion of new processes.
Given the large fractional saturation of the unitarity bounds by
$\lb\rightarrow\lc$, the inclusion of $\lb\rightarrow\lc^*$ could be
particularly fruitful once such data is available.

\begin{acknowledgments}
This work was supported by the U.S.\ Department of Energy under
Contract No.\ DE-FG02-93ER-40762 (T.D.C.\ and H.L.) and the National
Science Foundation under Grant No.\ PHY-1803912 (R.F.L.).  H.L.\
acknowledges the hospitality of Arizona State University, where part
of this work was performed.
\end{acknowledgments}

\bibliographystyle{apsrev4-1}
\bibliography{wise}

\begin{thebibliography}{71}%
\makeatletter
\providecommand \@ifxundefined [1]{%
 \@ifx{#1\undefined}
}%
\providecommand \@ifnum [1]{%
 \ifnum #1\expandafter \@firstoftwo
 \else \expandafter \@secondoftwo
 \fi
}%
\providecommand \@ifx [1]{%
 \ifx #1\expandafter \@firstoftwo
 \else \expandafter \@secondoftwo
 \fi
}%
\providecommand \natexlab [1]{#1}%
\providecommand \enquote  [1]{``#1''}%
\providecommand \bibnamefont  [1]{#1}%
\providecommand \bibfnamefont [1]{#1}%
\providecommand \citenamefont [1]{#1}%
\providecommand \href@noop [0]{\@secondoftwo}%
\providecommand \href [0]{\begingroup \@sanitize@url \@href}%
\providecommand \@href[1]{\@@startlink{#1}\@@href}%
\providecommand \@@href[1]{\endgroup#1\@@endlink}%
\providecommand \@sanitize@url [0]{\catcode `\\12\catcode `\$12\catcode
  `\&12\catcode `\#12\catcode `\^12\catcode `\_12\catcode `\%12\relax}%
\providecommand \@@startlink[1]{}%
\providecommand \@@endlink[0]{}%
\providecommand \url  [0]{\begingroup\@sanitize@url \@url }%
\providecommand \@url [1]{\endgroup\@href {#1}{\urlprefix }}%
\providecommand \urlprefix  [0]{URL }%
\providecommand \Eprint [0]{\href }%
\providecommand \doibase [0]{http://dx.doi.org/}%
\providecommand \selectlanguage [0]{\@gobble}%
\providecommand \bibinfo  [0]{\@secondoftwo}%
\providecommand \bibfield  [0]{\@secondoftwo}%
\providecommand \translation [1]{[#1]}%
\providecommand \BibitemOpen [0]{}%
\providecommand \bibitemStop [0]{}%
\providecommand \bibitemNoStop [0]{.\EOS\space}%
\providecommand \EOS [0]{\spacefactor3000\relax}%
\providecommand \BibitemShut  [1]{\csname bibitem#1\endcsname}%
\let\auto@bib@innerbib\@empty
\bibitem [{\citenamefont {Colangelo}\ and\ \citenamefont
  {De~Fazio}(2018)}]{Colangelo:2018cnj}%
  \BibitemOpen
  \bibfield  {author} {\bibinfo {author} {\bibfnamefont {P.}~\bibnamefont
  {Colangelo}}\ and\ \bibinfo {author} {\bibfnamefont {F.}~\bibnamefont
  {De~Fazio}},\ }\href {\doibase 10.1007/JHEP06(2018)082} {\bibfield  {journal}
  {\bibinfo  {journal} {JHEP}\ }\textbf {\bibinfo {volume} {06}},\ \bibinfo
  {pages} {082} (\bibinfo {year} {2018})},\ \Eprint
  {http://arxiv.org/abs/1801.10468} {arXiv:1801.10468 [hep-ph]} \BibitemShut
  {NoStop}%
\bibitem [{\citenamefont {Ivanov}\ \emph {et~al.}(2016)\citenamefont {Ivanov},
  \citenamefont {K{\" o}rner},\ and\ \citenamefont {Tran}}]{Ivanov:2016qtw}%
  \BibitemOpen
  \bibfield  {author} {\bibinfo {author} {\bibfnamefont {M.~A.}\ \bibnamefont
  {Ivanov}}, \bibinfo {author} {\bibfnamefont {J.~G.}\ \bibnamefont {K{\"
  o}rner}}, \ and\ \bibinfo {author} {\bibfnamefont {C.-T.}\ \bibnamefont
  {Tran}},\ }\href {\doibase 10.1103/PhysRevD.94.094028} {\bibfield  {journal}
  {\bibinfo  {journal} {Phys.\ Rev.}\ }\textbf {\bibinfo {volume} {D94}},\
  \bibinfo {pages} {094028} (\bibinfo {year} {2016})},\ \Eprint
  {http://arxiv.org/abs/1607.02932} {arXiv:1607.02932 [hep-ph]} \BibitemShut
  {NoStop}%
\bibitem [{\citenamefont {Tran}\ \emph {et~al.}(2018)\citenamefont {Tran},
  \citenamefont {Ivanov}, \citenamefont {K{\" o}rner},\ and\ \citenamefont
  {Santorelli}}]{Tran:2018kuv}%
  \BibitemOpen
  \bibfield  {author} {\bibinfo {author} {\bibfnamefont {C.-T.}\ \bibnamefont
  {Tran}}, \bibinfo {author} {\bibfnamefont {M.~A.}\ \bibnamefont {Ivanov}},
  \bibinfo {author} {\bibfnamefont {J.~G.}\ \bibnamefont {K{\" o}rner}}, \ and\
  \bibinfo {author} {\bibfnamefont {P.}~\bibnamefont {Santorelli}},\ }\href
  {\doibase 10.1103/PhysRevD.97.054014} {\bibfield  {journal} {\bibinfo
  {journal} {Phys.\ Rev.}\ }\textbf {\bibinfo {volume} {D97}},\ \bibinfo
  {pages} {054014} (\bibinfo {year} {2018})},\ \Eprint
  {http://arxiv.org/abs/1801.06927} {arXiv:1801.06927 [hep-ph]} \BibitemShut
  {NoStop}%
\bibitem [{\citenamefont {Bhattacharya}\ \emph {et~al.}(2016)\citenamefont
  {Bhattacharya}, \citenamefont {Nandi},\ and\ \citenamefont
  {Patra}}]{Bhattacharya:2015ida}%
  \BibitemOpen
  \bibfield  {author} {\bibinfo {author} {\bibfnamefont {S.}~\bibnamefont
  {Bhattacharya}}, \bibinfo {author} {\bibfnamefont {S.}~\bibnamefont {Nandi}},
  \ and\ \bibinfo {author} {\bibfnamefont {S.~K.}\ \bibnamefont {Patra}},\
  }\href {\doibase 10.1103/PhysRevD.93.034011} {\bibfield  {journal} {\bibinfo
  {journal} {Phys.\ Rev.}\ }\textbf {\bibinfo {volume} {D93}},\ \bibinfo
  {pages} {034011} (\bibinfo {year} {2016})},\ \Eprint
  {http://arxiv.org/abs/1509.07259} {arXiv:1509.07259 [hep-ph]} \BibitemShut
  {NoStop}%
\bibitem [{\citenamefont {Bhattacharya}\ \emph {et~al.}(2017)\citenamefont
  {Bhattacharya}, \citenamefont {Nandi},\ and\ \citenamefont
  {Patra}}]{Bhattacharya:2016zcw}%
  \BibitemOpen
  \bibfield  {author} {\bibinfo {author} {\bibfnamefont {S.}~\bibnamefont
  {Bhattacharya}}, \bibinfo {author} {\bibfnamefont {S.}~\bibnamefont {Nandi}},
  \ and\ \bibinfo {author} {\bibfnamefont {S.~K.}\ \bibnamefont {Patra}},\
  }\href {\doibase 10.1103/PhysRevD.95.075012} {\bibfield  {journal} {\bibinfo
  {journal} {Phys.\ Rev.}\ }\textbf {\bibinfo {volume} {D95}},\ \bibinfo
  {pages} {075012} (\bibinfo {year} {2017})},\ \Eprint
  {http://arxiv.org/abs/1611.04605} {arXiv:1611.04605 [hep-ph]} \BibitemShut
  {NoStop}%
\bibitem [{\citenamefont {Jaiswal}\ \emph {et~al.}(2017)\citenamefont
  {Jaiswal}, \citenamefont {Nandi},\ and\ \citenamefont
  {Patra}}]{Jaiswal:2017rve}%
  \BibitemOpen
  \bibfield  {author} {\bibinfo {author} {\bibfnamefont {S.}~\bibnamefont
  {Jaiswal}}, \bibinfo {author} {\bibfnamefont {S.}~\bibnamefont {Nandi}}, \
  and\ \bibinfo {author} {\bibfnamefont {S.~K.}\ \bibnamefont {Patra}},\ }\href
  {\doibase 10.1007/JHEP12(2017)060} {\bibfield  {journal} {\bibinfo  {journal}
  {JHEP}\ }\textbf {\bibinfo {volume} {12}},\ \bibinfo {pages} {060} (\bibinfo
  {year} {2017})},\ \Eprint {http://arxiv.org/abs/1707.09977} {arXiv:1707.09977
  [hep-ph]} \BibitemShut {NoStop}%
\bibitem [{\citenamefont {Bhattacharya}\ \emph {et~al.}(2019)\citenamefont
  {Bhattacharya}, \citenamefont {Nandi},\ and\ \citenamefont
  {Kumar~Patra}}]{Bhattacharya:2018kig}%
  \BibitemOpen
  \bibfield  {author} {\bibinfo {author} {\bibfnamefont {S.}~\bibnamefont
  {Bhattacharya}}, \bibinfo {author} {\bibfnamefont {S.}~\bibnamefont {Nandi}},
  \ and\ \bibinfo {author} {\bibfnamefont {S.}~\bibnamefont {Kumar~Patra}},\
  }\href {\doibase 10.1140/epjc/s10052-019-6767-7} {\bibfield  {journal}
  {\bibinfo  {journal} {Eur.\ Phys.\ J.}\ }\textbf {\bibinfo {volume} {C79}},\
  \bibinfo {pages} {268} (\bibinfo {year} {2019})},\ \Eprint
  {http://arxiv.org/abs/1805.08222} {arXiv:1805.08222 [hep-ph]} \BibitemShut
  {NoStop}%
\bibitem [{\citenamefont {Cohen}\ \emph
  {et~al.}(2018{\natexlab{a}})\citenamefont {Cohen}, \citenamefont {Lamm},\
  and\ \citenamefont {Lebed}}]{Cohen:2018vhw}%
  \BibitemOpen
  \bibfield  {author} {\bibinfo {author} {\bibfnamefont {T.~D.}\ \bibnamefont
  {Cohen}}, \bibinfo {author} {\bibfnamefont {H.}~\bibnamefont {Lamm}}, \ and\
  \bibinfo {author} {\bibfnamefont {R.~F.}\ \bibnamefont {Lebed}},\ }\href
  {\doibase 10.1103/PhysRevD.98.034022} {\bibfield  {journal} {\bibinfo
  {journal} {Phys.\ Rev.}\ }\textbf {\bibinfo {volume} {D98}},\ \bibinfo
  {pages} {034022} (\bibinfo {year} {2018}{\natexlab{a}})},\ \Eprint
  {http://arxiv.org/abs/1807.00256} {arXiv:1807.00256 [hep-ph]} \BibitemShut
  {NoStop}%
\bibitem [{\citenamefont {Be\v{c}irevi\'{c}}\ \emph {et~al.}(2019)\citenamefont
  {Be\v{c}irevi\'{c}}, \citenamefont {Fedele}, \citenamefont
  {Ni\v{s}and\v{z}i\'{c}},\ and\ \citenamefont
  {Tayduganov}}]{Becirevic:2019tpx}%
  \BibitemOpen
  \bibfield  {author} {\bibinfo {author} {\bibfnamefont {D.}~\bibnamefont
  {Be\v{c}irevi\'{c}}}, \bibinfo {author} {\bibfnamefont {M.}~\bibnamefont
  {Fedele}}, \bibinfo {author} {\bibfnamefont {I.}~\bibnamefont
  {Ni\v{s}and\v{z}i\'{c}}}, \ and\ \bibinfo {author} {\bibfnamefont
  {A.}~\bibnamefont {Tayduganov}},\ }\href@noop {} {\  (\bibinfo {year}
  {2019})},\ \Eprint {http://arxiv.org/abs/1907.02257} {arXiv:1907.02257
  [hep-ph]} \BibitemShut {NoStop}%
\bibitem [{\citenamefont {Lees}\ \emph {et~al.}(2012)\citenamefont {Lees} \emph
  {et~al.}}]{Lees:2012xj}%
  \BibitemOpen
  \bibfield  {author} {\bibinfo {author} {\bibfnamefont {J.~P.}\ \bibnamefont
  {Lees}} \emph {et~al.} (\bibinfo {collaboration} {BaBar Collaboration}),\
  }\href {\doibase 10.1103/PhysRevLett.109.101802} {\bibfield  {journal}
  {\bibinfo  {journal} {Phys.\ Rev.\ Lett.}\ }\textbf {\bibinfo {volume}
  {109}},\ \bibinfo {pages} {101802} (\bibinfo {year} {2012})},\ \Eprint
  {http://arxiv.org/abs/1205.5442} {arXiv:1205.5442 [hep-ex]} \BibitemShut
  {NoStop}%
\bibitem [{\citenamefont {Lees}\ \emph {et~al.}(2013)\citenamefont {Lees} \emph
  {et~al.}}]{Lees:2013uzd}%
  \BibitemOpen
  \bibfield  {author} {\bibinfo {author} {\bibfnamefont {J.~P.}\ \bibnamefont
  {Lees}} \emph {et~al.} (\bibinfo {collaboration} {BaBar Collaboration}),\
  }\href {\doibase 10.1103/PhysRevD.88.072012} {\bibfield  {journal} {\bibinfo
  {journal} {Phys.\ Rev.}\ }\textbf {\bibinfo {volume} {D88}},\ \bibinfo
  {pages} {072012} (\bibinfo {year} {2013})},\ \Eprint
  {http://arxiv.org/abs/1303.0571} {arXiv:1303.0571 [hep-ex]} \BibitemShut
  {NoStop}%
\bibitem [{\citenamefont {Huschle}\ \emph {et~al.}(2015)\citenamefont {Huschle}
  \emph {et~al.}}]{Huschle:2015rga}%
  \BibitemOpen
  \bibfield  {author} {\bibinfo {author} {\bibfnamefont {M.}~\bibnamefont
  {Huschle}} \emph {et~al.} (\bibinfo {collaboration} {Belle Collaboration}),\
  }\href {\doibase 10.1103/PhysRevD.92.072014} {\bibfield  {journal} {\bibinfo
  {journal} {Phys.\ Rev.}\ }\textbf {\bibinfo {volume} {D92}},\ \bibinfo
  {pages} {072014} (\bibinfo {year} {2015})},\ \Eprint
  {http://arxiv.org/abs/1507.03233} {arXiv:1507.03233 [hep-ex]} \BibitemShut
  {NoStop}%
\bibitem [{\citenamefont {Abdesselam}\ \emph
  {et~al.}(2019{\natexlab{a}})\citenamefont {Abdesselam} \emph
  {et~al.}}]{Abdesselam:2019dgh}%
  \BibitemOpen
  \bibfield  {author} {\bibinfo {author} {\bibfnamefont {A.}~\bibnamefont
  {Abdesselam}} \emph {et~al.} (\bibinfo {collaboration} {Belle
  Collaboration}),\ }\href@noop {} {\  (\bibinfo {year}
  {2019}{\natexlab{a}})},\ \Eprint {http://arxiv.org/abs/1904.08794}
  {arXiv:1904.08794 [hep-ex]} \BibitemShut {NoStop}%
\bibitem [{\citenamefont {Sato}\ \emph {et~al.}(2016)\citenamefont {Sato} \emph
  {et~al.}}]{Sato:2016svk}%
  \BibitemOpen
  \bibfield  {author} {\bibinfo {author} {\bibfnamefont {Y.}~\bibnamefont
  {Sato}} \emph {et~al.} (\bibinfo {collaboration} {Belle Collaboration}),\
  }\href {\doibase 10.1103/PhysRevD.94.072007} {\bibfield  {journal} {\bibinfo
  {journal} {Phys.\ Rev.}\ }\textbf {\bibinfo {volume} {D94}},\ \bibinfo
  {pages} {072007} (\bibinfo {year} {2016})},\ \Eprint
  {http://arxiv.org/abs/1607.07923} {arXiv:1607.07923 [hep-ex]} \BibitemShut
  {NoStop}%
\bibitem [{\citenamefont {Aaij}\ \emph {et~al.}(2015)\citenamefont {Aaij} \emph
  {et~al.}}]{Aaij:2015yra}%
  \BibitemOpen
  \bibfield  {author} {\bibinfo {author} {\bibfnamefont {R.}~\bibnamefont
  {Aaij}} \emph {et~al.} (\bibinfo {collaboration} {LHCb Collaboration}),\
  }\href {\doibase 10.1103/PhysRevLett.115.159901,
  10.1103/PhysRevLett.115.111803} {\bibfield  {journal} {\bibinfo  {journal}
  {Phys.\ Rev.\ Lett.}\ }\textbf {\bibinfo {volume} {115}},\ \bibinfo {pages}
  {111803} (\bibinfo {year} {2015})},\ \bibinfo {note} {[Erratum: Phys.\ Rev.\
  Lett.\ {\bf 115}, 159901 (2015)]},\ \Eprint {http://arxiv.org/abs/1506.08614}
  {arXiv:1506.08614 [hep-ex]} \BibitemShut {NoStop}%
\bibitem [{\citenamefont {Hirose}\ \emph {et~al.}(2017)\citenamefont {Hirose}
  \emph {et~al.}}]{Hirose:2016wfn}%
  \BibitemOpen
  \bibfield  {author} {\bibinfo {author} {\bibfnamefont {S.}~\bibnamefont
  {Hirose}} \emph {et~al.} (\bibinfo {collaboration} {Belle Collaboration}),\
  }\href {\doibase 10.1103/PhysRevLett.118.211801} {\bibfield  {journal}
  {\bibinfo  {journal} {Phys.\ Rev.\ Lett.}\ }\textbf {\bibinfo {volume}
  {118}},\ \bibinfo {pages} {211801} (\bibinfo {year} {2017})},\ \Eprint
  {http://arxiv.org/abs/1612.00529} {arXiv:1612.00529 [hep-ex]} \BibitemShut
  {NoStop}%
\bibitem [{\citenamefont {Aaij}\ \emph
  {et~al.}(2018{\natexlab{a}})\citenamefont {Aaij} \emph
  {et~al.}}]{Aaij:2017uff}%
  \BibitemOpen
  \bibfield  {author} {\bibinfo {author} {\bibfnamefont {R.}~\bibnamefont
  {Aaij}} \emph {et~al.} (\bibinfo {collaboration} {LHCb Collaboration}),\
  }\href {\doibase 10.1103/PhysRevLett.120.171802} {\bibfield  {journal}
  {\bibinfo  {journal} {Phys.\ Rev.\ Lett.}\ }\textbf {\bibinfo {volume}
  {120}},\ \bibinfo {pages} {171802} (\bibinfo {year} {2018}{\natexlab{a}})},\
  \Eprint {http://arxiv.org/abs/1708.08856} {arXiv:1708.08856 [hep-ex]}
  \BibitemShut {NoStop}%
\bibitem [{\citenamefont {Aaij}\ \emph
  {et~al.}(2018{\natexlab{b}})\citenamefont {Aaij} \emph
  {et~al.}}]{Aaij:2017deq}%
  \BibitemOpen
  \bibfield  {author} {\bibinfo {author} {\bibfnamefont {R.}~\bibnamefont
  {Aaij}} \emph {et~al.} (\bibinfo {collaboration} {LHCb Collaboration}),\
  }\href {\doibase 10.1103/PhysRevD.97.072013} {\bibfield  {journal} {\bibinfo
  {journal} {Phys. Rev.}\ }\textbf {\bibinfo {volume} {D97}},\ \bibinfo {pages}
  {072013} (\bibinfo {year} {2018}{\natexlab{b}})},\ \Eprint
  {http://arxiv.org/abs/1711.02505} {arXiv:1711.02505 [hep-ex]} \BibitemShut
  {NoStop}%
\bibitem [{\citenamefont {Hirose}\ \emph {et~al.}(2018)\citenamefont {Hirose}
  \emph {et~al.}}]{Hirose:2017dxl}%
  \BibitemOpen
  \bibfield  {author} {\bibinfo {author} {\bibfnamefont {S.}~\bibnamefont
  {Hirose}} \emph {et~al.} (\bibinfo {collaboration} {Belle Collaboration}),\
  }\href {\doibase 10.1103/PhysRevD.97.012004} {\bibfield  {journal} {\bibinfo
  {journal} {Phys.\ Rev.}\ }\textbf {\bibinfo {volume} {D97}},\ \bibinfo
  {pages} {012004} (\bibinfo {year} {2018})},\ \Eprint
  {http://arxiv.org/abs/1709.00129} {arXiv:1709.00129 [hep-ex]} \BibitemShut
  {NoStop}%
\bibitem [{\citenamefont {Amhis}\ \emph {et~al.}(2017)\citenamefont {Amhis}
  \emph {et~al.}}]{Amhis:2016xyh}%
  \BibitemOpen
  \bibfield  {author} {\bibinfo {author} {\bibfnamefont {Y.}~\bibnamefont
  {Amhis}} \emph {et~al.} (\bibinfo {collaboration} {Heavy Flavor Averaging
  Group}),\ }\href {\doibase 10.1140/epjc/s10052-017-5058-4} {\bibfield
  {journal} {\bibinfo  {journal} {Eur.\ Phys.\ J.}\ }\textbf {\bibinfo {volume}
  {C77}},\ \bibinfo {pages} {895} (\bibinfo {year} {2017})},\ \bibinfo {note}
  {{updated results and plots available at
  \href{https://hflav.web.cern.ch}{{\texttt{https://hflav.web.cern.ch}}}}},\
  \Eprint {http://arxiv.org/abs/1612.07233} {arXiv:1612.07233 [hep-ex]}
  \BibitemShut {NoStop}%
\bibitem [{\citenamefont {Bigi}\ and\ \citenamefont
  {Gambino}(2016)}]{Bigi:2016mdz}%
  \BibitemOpen
  \bibfield  {author} {\bibinfo {author} {\bibfnamefont {D.}~\bibnamefont
  {Bigi}}\ and\ \bibinfo {author} {\bibfnamefont {P.}~\bibnamefont {Gambino}},\
  }\href {\doibase 10.1103/PhysRevD.94.094008} {\bibfield  {journal} {\bibinfo
  {journal} {Phys.\ Rev.}\ }\textbf {\bibinfo {volume} {D94}},\ \bibinfo
  {pages} {094008} (\bibinfo {year} {2016})},\ \Eprint
  {http://arxiv.org/abs/1606.08030} {arXiv:1606.08030 [hep-ph]} \BibitemShut
  {NoStop}%
\bibitem [{\citenamefont {Bernlochner}\ \emph {et~al.}(2017)\citenamefont
  {Bernlochner}, \citenamefont {Ligeti}, \citenamefont {Papucci},\ and\
  \citenamefont {Robinson}}]{Bernlochner:2017jka}%
  \BibitemOpen
  \bibfield  {author} {\bibinfo {author} {\bibfnamefont {F.~U.}\ \bibnamefont
  {Bernlochner}}, \bibinfo {author} {\bibfnamefont {Z.}~\bibnamefont {Ligeti}},
  \bibinfo {author} {\bibfnamefont {M.}~\bibnamefont {Papucci}}, \ and\
  \bibinfo {author} {\bibfnamefont {D.~J.}\ \bibnamefont {Robinson}},\ }\href
  {\doibase 10.1103/PhysRevD.95.115008, 10.1103/PhysRevD.97.059902} {\bibfield
  {journal} {\bibinfo  {journal} {Phys.\ Rev.}\ }\textbf {\bibinfo {volume}
  {D95}},\ \bibinfo {pages} {115008} (\bibinfo {year} {2017})},\ \bibinfo
  {note} {[Erratum: Phys.\ Rev. {\bf D97}, 059902 (2018)]},\ \Eprint
  {http://arxiv.org/abs/1703.05330} {arXiv:1703.05330 [hep-ph]} \BibitemShut
  {NoStop}%
\bibitem [{\citenamefont {Bigi}\ \emph {et~al.}(2017)\citenamefont {Bigi},
  \citenamefont {Gambino},\ and\ \citenamefont {Schacht}}]{Bigi:2017jbd}%
  \BibitemOpen
  \bibfield  {author} {\bibinfo {author} {\bibfnamefont {D.}~\bibnamefont
  {Bigi}}, \bibinfo {author} {\bibfnamefont {P.}~\bibnamefont {Gambino}}, \
  and\ \bibinfo {author} {\bibfnamefont {S.}~\bibnamefont {Schacht}},\ }\href
  {\doibase 10.1007/JHEP11(2017)061} {\bibfield  {journal} {\bibinfo  {journal}
  {JHEP}\ }\textbf {\bibinfo {volume} {11}},\ \bibinfo {pages} {061} (\bibinfo
  {year} {2017})},\ \Eprint {http://arxiv.org/abs/1707.09509} {arXiv:1707.09509
  [hep-ph]} \BibitemShut {NoStop}%
\bibitem [{\citenamefont {Aaij}\ \emph
  {et~al.}(2018{\natexlab{c}})\citenamefont {Aaij} \emph
  {et~al.}}]{Aaij:2017tyk}%
  \BibitemOpen
  \bibfield  {author} {\bibinfo {author} {\bibfnamefont {R.}~\bibnamefont
  {Aaij}} \emph {et~al.} (\bibinfo {collaboration} {LHCb Collaboration}),\
  }\href {\doibase 10.1103/PhysRevLett.120.121801} {\bibfield  {journal}
  {\bibinfo  {journal} {Phys.\ Rev.\ Lett.}\ }\textbf {\bibinfo {volume}
  {120}},\ \bibinfo {pages} {121801} (\bibinfo {year} {2018}{\natexlab{c}})},\
  \Eprint {http://arxiv.org/abs/1711.05623} {arXiv:1711.05623 [hep-ex]}
  \BibitemShut {NoStop}%
\bibitem [{\citenamefont {Cohen}\ \emph
  {et~al.}(2018{\natexlab{b}})\citenamefont {Cohen}, \citenamefont {Lamm},\
  and\ \citenamefont {Lebed}}]{Cohen:2018dgz}%
  \BibitemOpen
  \bibfield  {author} {\bibinfo {author} {\bibfnamefont {T.~D.}\ \bibnamefont
  {Cohen}}, \bibinfo {author} {\bibfnamefont {H.}~\bibnamefont {Lamm}}, \ and\
  \bibinfo {author} {\bibfnamefont {R.~F.}\ \bibnamefont {Lebed}},\ }\href
  {\doibase 10.1007/JHEP09(2018)168} {\bibfield  {journal} {\bibinfo  {journal}
  {JHEP}\ }\textbf {\bibinfo {volume} {09}},\ \bibinfo {pages} {168} (\bibinfo
  {year} {2018}{\natexlab{b}})},\ \Eprint {http://arxiv.org/abs/1807.02730}
  {arXiv:1807.02730 [hep-ph]} \BibitemShut {NoStop}%
\bibitem [{\citenamefont {Aoki}\ \emph {et~al.}(2019)\citenamefont {Aoki} \emph
  {et~al.}}]{Aoki:2019cca}%
  \BibitemOpen
  \bibfield  {author} {\bibinfo {author} {\bibfnamefont {S.}~\bibnamefont
  {Aoki}} \emph {et~al.} (\bibinfo {collaboration} {Flavour Lattice Averaging
  Group}),\ }\href@noop {} {\  (\bibinfo {year} {2019})},\ \Eprint
  {http://arxiv.org/abs/1902.08191} {arXiv:1902.08191 [hep-lat]} \BibitemShut
  {NoStop}%
\bibitem [{\citenamefont {Na}\ \emph {et~al.}(2015)\citenamefont {Na},
  \citenamefont {Bouchard}, \citenamefont {Lepage}, \citenamefont {Monahan},\
  and\ \citenamefont {Shigemitsu}}]{Na:2015kha}%
  \BibitemOpen
  \bibfield  {author} {\bibinfo {author} {\bibfnamefont {H.}~\bibnamefont
  {Na}}, \bibinfo {author} {\bibfnamefont {C.~M.}\ \bibnamefont {Bouchard}},
  \bibinfo {author} {\bibfnamefont {G.~P.}\ \bibnamefont {Lepage}}, \bibinfo
  {author} {\bibfnamefont {C.}~\bibnamefont {Monahan}}, \ and\ \bibinfo
  {author} {\bibfnamefont {J.}~\bibnamefont {Shigemitsu}} (\bibinfo
  {collaboration} {HPQCD Collaboration}),\ }\href {\doibase
  10.1103/PhysRevD.92.054510} {\bibfield  {journal} {\bibinfo  {journal}
  {Phys.\ Rev.}\ }\textbf {\bibinfo {volume} {D92}},\ \bibinfo {pages} {054510}
  (\bibinfo {year} {2015})},\ \bibinfo {note} {[Erratum: Phys.\ Rev.\ {\bf
  D93}, 119906 (2016)]},\ \Eprint {http://arxiv.org/abs/1505.03925}
  {arXiv:1505.03925 [hep-lat]} \BibitemShut {NoStop}%
\bibitem [{\citenamefont {Bailey}\ \emph {et~al.}(2015)\citenamefont {Bailey}
  \emph {et~al.}}]{Lattice:2015rga}%
  \BibitemOpen
  \bibfield  {author} {\bibinfo {author} {\bibfnamefont {J.~A.}\ \bibnamefont
  {Bailey}} \emph {et~al.} (\bibinfo {collaboration} {MILC Collaboration}),\
  }\href {\doibase 10.1103/PhysRevD.92.034506} {\bibfield  {journal} {\bibinfo
  {journal} {Phys.\ Rev.}\ }\textbf {\bibinfo {volume} {D92}},\ \bibinfo
  {pages} {034506} (\bibinfo {year} {2015})},\ \Eprint
  {http://arxiv.org/abs/1503.07237} {arXiv:1503.07237 [hep-lat]} \BibitemShut
  {NoStop}%
\bibitem [{\citenamefont {McLean}\ \emph
  {et~al.}(2019{\natexlab{a}})\citenamefont {McLean}, \citenamefont {Davies},
  \citenamefont {Koponen},\ and\ \citenamefont {Lytle}}]{McLean:2019qcx}%
  \BibitemOpen
  \bibfield  {author} {\bibinfo {author} {\bibfnamefont {E.}~\bibnamefont
  {McLean}}, \bibinfo {author} {\bibfnamefont {C.~T.~H.}\ \bibnamefont
  {Davies}}, \bibinfo {author} {\bibfnamefont {J.}~\bibnamefont {Koponen}}, \
  and\ \bibinfo {author} {\bibfnamefont {A.~T.}\ \bibnamefont {Lytle}},\
  }\href@noop {} {\  (\bibinfo {year} {2019}{\natexlab{a}})},\ \Eprint
  {http://arxiv.org/abs/1906.00701} {arXiv:1906.00701 [hep-lat]} \BibitemShut
  {NoStop}%
\bibitem [{\citenamefont {Detmold}\ \emph {et~al.}(2015)\citenamefont
  {Detmold}, \citenamefont {Lehner},\ and\ \citenamefont
  {Meinel}}]{Detmold:2015aaa}%
  \BibitemOpen
  \bibfield  {author} {\bibinfo {author} {\bibfnamefont {W.}~\bibnamefont
  {Detmold}}, \bibinfo {author} {\bibfnamefont {C.}~\bibnamefont {Lehner}}, \
  and\ \bibinfo {author} {\bibfnamefont {S.}~\bibnamefont {Meinel}},\ }\href
  {\doibase 10.1103/PhysRevD.92.034503} {\bibfield  {journal} {\bibinfo
  {journal} {Phys.\ Rev.}\ }\textbf {\bibinfo {volume} {D92}},\ \bibinfo
  {pages} {034503} (\bibinfo {year} {2015})},\ \Eprint
  {http://arxiv.org/abs/1503.01421} {arXiv:1503.01421 [hep-lat]} \BibitemShut
  {NoStop}%
\bibitem [{\citenamefont {Bernlochner}\ \emph {et~al.}(2018)\citenamefont
  {Bernlochner}, \citenamefont {Ligeti}, \citenamefont {Robinson},\ and\
  \citenamefont {Sutcliffe}}]{Bernlochner:2018kxh}%
  \BibitemOpen
  \bibfield  {author} {\bibinfo {author} {\bibfnamefont {F.~U.}\ \bibnamefont
  {Bernlochner}}, \bibinfo {author} {\bibfnamefont {Z.}~\bibnamefont {Ligeti}},
  \bibinfo {author} {\bibfnamefont {D.~J.}\ \bibnamefont {Robinson}}, \ and\
  \bibinfo {author} {\bibfnamefont {W.~L.}\ \bibnamefont {Sutcliffe}},\ }\href
  {\doibase 10.1103/PhysRevLett.121.202001} {\bibfield  {journal} {\bibinfo
  {journal} {Phys.\ Rev.\ Lett.}\ }\textbf {\bibinfo {volume} {121}},\ \bibinfo
  {pages} {202001} (\bibinfo {year} {2018})},\ \Eprint
  {http://arxiv.org/abs/1808.09464} {arXiv:1808.09464 [hep-ph]} \BibitemShut
  {NoStop}%
\bibitem [{\citenamefont {Berns}\ and\ \citenamefont
  {Lamm}(2018)}]{Berns:2018vpl}%
  \BibitemOpen
  \bibfield  {author} {\bibinfo {author} {\bibfnamefont {A.}~\bibnamefont
  {Berns}}\ and\ \bibinfo {author} {\bibfnamefont {H.}~\bibnamefont {Lamm}},\
  }\href {\doibase 10.1007/JHEP12(2018)114} {\bibfield  {journal} {\bibinfo
  {journal} {JHEP}\ }\textbf {\bibinfo {volume} {12}},\ \bibinfo {pages} {114}
  (\bibinfo {year} {2018})},\ \Eprint {http://arxiv.org/abs/1808.07360}
  {arXiv:1808.07360 [hep-ph]} \BibitemShut {NoStop}%
\bibitem [{\citenamefont {Murphy}\ and\ \citenamefont
  {Soni}(2018)}]{Murphy:2018sqg}%
  \BibitemOpen
  \bibfield  {author} {\bibinfo {author} {\bibfnamefont {C.~W.}\ \bibnamefont
  {Murphy}}\ and\ \bibinfo {author} {\bibfnamefont {A.}~\bibnamefont {Soni}},\
  }\href {\doibase 10.1103/PhysRevD.98.094026} {\bibfield  {journal} {\bibinfo
  {journal} {Phys.\ Rev.}\ }\textbf {\bibinfo {volume} {D98}},\ \bibinfo
  {pages} {094026} (\bibinfo {year} {2018})},\ \Eprint
  {http://arxiv.org/abs/1808.05932} {arXiv:1808.05932 [hep-ph]} \BibitemShut
  {NoStop}%
\bibitem [{\citenamefont {Cerri}\ \emph {et~al.}(2018)\citenamefont {Cerri}
  \emph {et~al.}}]{Cerri:2018ypt}%
  \BibitemOpen
  \bibfield  {author} {\bibinfo {author} {\bibfnamefont {A.}~\bibnamefont
  {Cerri}} \emph {et~al.},\ }\href@noop {} {\  (\bibinfo {year} {2018})},\
  \Eprint {http://arxiv.org/abs/1812.07638} {arXiv:1812.07638 [hep-ph]}
  \BibitemShut {NoStop}%
\bibitem [{\citenamefont {Hamilton}\ and\ \citenamefont {Jawahery}()}]{BHHJ}%
  \BibitemOpen
  \bibfield  {author} {\bibinfo {author} {\bibfnamefont {B.}~\bibnamefont
  {Hamilton}}\ and\ \bibinfo {author} {\bibfnamefont {H.}~\bibnamefont
  {Jawahery}},\ }\href@noop {} {}\bibinfo {howpublished} {private
  communication}\BibitemShut {NoStop}%
\bibitem [{\citenamefont {Boyd}\ \emph
  {et~al.}(1995{\natexlab{a}})\citenamefont {Boyd}, \citenamefont {Grinstein},\
  and\ \citenamefont {Lebed}}]{Boyd:1995cf}%
  \BibitemOpen
  \bibfield  {author} {\bibinfo {author} {\bibfnamefont {C.~G.}\ \bibnamefont
  {Boyd}}, \bibinfo {author} {\bibfnamefont {B.}~\bibnamefont {Grinstein}}, \
  and\ \bibinfo {author} {\bibfnamefont {R.~F.}\ \bibnamefont {Lebed}},\ }\href
  {\doibase 10.1016/0370-2693(95)00480-9} {\bibfield  {journal} {\bibinfo
  {journal} {Phys.\ Lett.}\ }\textbf {\bibinfo {volume} {B353}},\ \bibinfo
  {pages} {306} (\bibinfo {year} {1995}{\natexlab{a}})},\ \Eprint
  {http://arxiv.org/abs/hep-ph/9504235} {arXiv:hep-ph/9504235 [hep-ph]}
  \BibitemShut {NoStop}%
\bibitem [{\citenamefont {Boyd}\ \emph {et~al.}(1997)\citenamefont {Boyd},
  \citenamefont {Grinstein},\ and\ \citenamefont {Lebed}}]{Boyd:1997kz}%
  \BibitemOpen
  \bibfield  {author} {\bibinfo {author} {\bibfnamefont {C.~G.}\ \bibnamefont
  {Boyd}}, \bibinfo {author} {\bibfnamefont {B.}~\bibnamefont {Grinstein}}, \
  and\ \bibinfo {author} {\bibfnamefont {R.~F.}\ \bibnamefont {Lebed}},\ }\href
  {\doibase 10.1103/PhysRevD.56.6895} {\bibfield  {journal} {\bibinfo
  {journal} {Phys.\ Rev.}\ }\textbf {\bibinfo {volume} {D56}},\ \bibinfo
  {pages} {6895} (\bibinfo {year} {1997})},\ \Eprint
  {http://arxiv.org/abs/hep-ph/9705252} {arXiv:hep-ph/9705252 [hep-ph]}
  \BibitemShut {NoStop}%
\bibitem [{\citenamefont {Grinstein}\ and\ \citenamefont
  {Lebed}(2015)}]{Grinstein:2015wqa}%
  \BibitemOpen
  \bibfield  {author} {\bibinfo {author} {\bibfnamefont {B.}~\bibnamefont
  {Grinstein}}\ and\ \bibinfo {author} {\bibfnamefont {R.~F.}\ \bibnamefont
  {Lebed}},\ }\href {\doibase 10.1103/PhysRevD.92.116001} {\bibfield  {journal}
  {\bibinfo  {journal} {Phys.\ Rev.}\ }\textbf {\bibinfo {volume} {D92}},\
  \bibinfo {pages} {116001} (\bibinfo {year} {2015})},\ \Eprint
  {http://arxiv.org/abs/1509.04847} {arXiv:1509.04847 [hep-ph]} \BibitemShut
  {NoStop}%
\bibitem [{\citenamefont {Colquhoun}\ \emph {et~al.}(2016)\citenamefont
  {Colquhoun}, \citenamefont {Davies}, \citenamefont {Koponen}, \citenamefont
  {Lytle},\ and\ \citenamefont {McNeile}}]{Colquhoun:2016osw}%
  \BibitemOpen
  \bibfield  {author} {\bibinfo {author} {\bibfnamefont {B.}~\bibnamefont
  {Colquhoun}}, \bibinfo {author} {\bibfnamefont {C.}~\bibnamefont {Davies}},
  \bibinfo {author} {\bibfnamefont {J.}~\bibnamefont {Koponen}}, \bibinfo
  {author} {\bibfnamefont {A.}~\bibnamefont {Lytle}}, \ and\ \bibinfo {author}
  {\bibfnamefont {C.}~\bibnamefont {McNeile}} (\bibinfo {collaboration} {HPQCD
  Collaboration}),\ }\bibfield  {booktitle} {\emph {\bibinfo {booktitle}
  {{Proceedings, 34th International Symposium on Lattice Field Theory (Lattice
  2016): Southampton, UK, July 24--30, 2016}}},\ }\href@noop {} {\bibfield
  {journal} {\bibinfo  {journal} {PoS}\ }\textbf {\bibinfo {volume} {LATTICE
  2016}},\ \bibinfo {pages} {281} (\bibinfo {year} {2016})},\ \Eprint
  {http://arxiv.org/abs/1611.01987} {arXiv:1611.01987 [hep-lat]} \BibitemShut
  {NoStop}%
\bibitem [{\citenamefont {Lytle}()}]{ALE}%
  \BibitemOpen
  \bibfield  {author} {\bibinfo {author} {\bibfnamefont {A.}~\bibnamefont
  {Lytle}},\ }\href@noop {} {}\bibinfo {howpublished} {private
  communication}\BibitemShut {NoStop}%
\bibitem [{\citenamefont {Wirbel}\ \emph {et~al.}(1985)\citenamefont {Wirbel},
  \citenamefont {Stech},\ and\ \citenamefont {Bauer}}]{Wirbel:1985ji}%
  \BibitemOpen
  \bibfield  {author} {\bibinfo {author} {\bibfnamefont {M.}~\bibnamefont
  {Wirbel}}, \bibinfo {author} {\bibfnamefont {B.}~\bibnamefont {Stech}}, \
  and\ \bibinfo {author} {\bibfnamefont {M.}~\bibnamefont {Bauer}},\ }\href
  {\doibase 10.1007/BF01560299} {\bibfield  {journal} {\bibinfo  {journal} {Z.
  Phys.}\ }\textbf {\bibinfo {volume} {C29}},\ \bibinfo {pages} {637} (\bibinfo
  {year} {1985})}\BibitemShut {NoStop}%
\bibitem [{\citenamefont {Richman}\ and\ \citenamefont
  {Burchat}(1995)}]{Richman:1995wm}%
  \BibitemOpen
  \bibfield  {author} {\bibinfo {author} {\bibfnamefont {J.~D.}\ \bibnamefont
  {Richman}}\ and\ \bibinfo {author} {\bibfnamefont {P.~R.}\ \bibnamefont
  {Burchat}},\ }\href {\doibase 10.1103/RevModPhys.67.893} {\bibfield
  {journal} {\bibinfo  {journal} {Rev.\ Mod.\ Phys.}\ }\textbf {\bibinfo
  {volume} {67}},\ \bibinfo {pages} {893} (\bibinfo {year} {1995})},\ \Eprint
  {http://arxiv.org/abs/hep-ph/9508250} {arXiv:hep-ph/9508250 [hep-ph]}
  \BibitemShut {NoStop}%
\bibitem [{\citenamefont {Isgur}\ and\ \citenamefont
  {Wise}(1989)}]{Isgur:1989vq}%
  \BibitemOpen
  \bibfield  {author} {\bibinfo {author} {\bibfnamefont {N.}~\bibnamefont
  {Isgur}}\ and\ \bibinfo {author} {\bibfnamefont {M.~B.}\ \bibnamefont
  {Wise}},\ }\href {\doibase 10.1016/0370-2693(89)90566-2} {\bibfield
  {journal} {\bibinfo  {journal} {Phys.\ Lett.}\ }\textbf {\bibinfo {volume}
  {B232}},\ \bibinfo {pages} {113} (\bibinfo {year} {1989})}\BibitemShut
  {NoStop}%
\bibitem [{\citenamefont {Isgur}\ and\ \citenamefont
  {Wise}(1990)}]{Isgur:1989ed}%
  \BibitemOpen
  \bibfield  {author} {\bibinfo {author} {\bibfnamefont {N.}~\bibnamefont
  {Isgur}}\ and\ \bibinfo {author} {\bibfnamefont {M.~B.}\ \bibnamefont
  {Wise}},\ }\href {\doibase 10.1016/0370-2693(90)91219-2} {\bibfield
  {journal} {\bibinfo  {journal} {Phys.\ Lett.}\ }\textbf {\bibinfo {volume}
  {B237}},\ \bibinfo {pages} {527} (\bibinfo {year} {1990})}\BibitemShut
  {NoStop}%
\bibitem [{\citenamefont {Mannel}\ \emph {et~al.}(1991)\citenamefont {Mannel},
  \citenamefont {Roberts},\ and\ \citenamefont {Ryzak}}]{Mannel:1990vg}%
  \BibitemOpen
  \bibfield  {author} {\bibinfo {author} {\bibfnamefont {T.}~\bibnamefont
  {Mannel}}, \bibinfo {author} {\bibfnamefont {W.}~\bibnamefont {Roberts}}, \
  and\ \bibinfo {author} {\bibfnamefont {Z.}~\bibnamefont {Ryzak}},\ }\href
  {\doibase 10.1016/0550-3213(91)90301-D} {\bibfield  {journal} {\bibinfo
  {journal} {Nucl.\ Phys.}\ }\textbf {\bibinfo {volume} {B355}},\ \bibinfo
  {pages} {38} (\bibinfo {year} {1991})}\BibitemShut {NoStop}%
\bibitem [{\citenamefont {Jenkins}\ \emph {et~al.}(1993)\citenamefont
  {Jenkins}, \citenamefont {Luke}, \citenamefont {Manohar},\ and\ \citenamefont
  {Savage}}]{Jenkins:1992nb}%
  \BibitemOpen
  \bibfield  {author} {\bibinfo {author} {\bibfnamefont {E.~E.}\ \bibnamefont
  {Jenkins}}, \bibinfo {author} {\bibfnamefont {M.~E.}\ \bibnamefont {Luke}},
  \bibinfo {author} {\bibfnamefont {A.~V.}\ \bibnamefont {Manohar}}, \ and\
  \bibinfo {author} {\bibfnamefont {M.~J.}\ \bibnamefont {Savage}},\ }\href
  {\doibase 10.1016/0550-3213(93)90464-Z} {\bibfield  {journal} {\bibinfo
  {journal} {Nucl.\ Phys.}\ }\textbf {\bibinfo {volume} {B390}},\ \bibinfo
  {pages} {463} (\bibinfo {year} {1993})},\ \Eprint
  {http://arxiv.org/abs/hep-ph/9204238} {arXiv:hep-ph/9204238 [hep-ph]}
  \BibitemShut {NoStop}%
\bibitem [{\citenamefont {Kiselev}\ \emph {et~al.}(2000)\citenamefont
  {Kiselev}, \citenamefont {Likhoded},\ and\ \citenamefont
  {Onishchenko}}]{Kiselev:1999sc}%
  \BibitemOpen
  \bibfield  {author} {\bibinfo {author} {\bibfnamefont {V.~V.}\ \bibnamefont
  {Kiselev}}, \bibinfo {author} {\bibfnamefont {A.~K.}\ \bibnamefont
  {Likhoded}}, \ and\ \bibinfo {author} {\bibfnamefont {A.~I.}\ \bibnamefont
  {Onishchenko}},\ }\href {\doibase 10.1016/S0550-3213(99)00505-2} {\bibfield
  {journal} {\bibinfo  {journal} {Nucl.\ Phys.}\ }\textbf {\bibinfo {volume}
  {B569}},\ \bibinfo {pages} {473} (\bibinfo {year} {2000})},\ \Eprint
  {http://arxiv.org/abs/hep-ph/9905359} {arXiv:hep-ph/9905359 [hep-ph]}
  \BibitemShut {NoStop}%
\bibitem [{\citenamefont {Falk}\ \emph {et~al.}(1990)\citenamefont {Falk},
  \citenamefont {Georgi}, \citenamefont {Grinstein},\ and\ \citenamefont
  {Wise}}]{Falk:1990yz}%
  \BibitemOpen
  \bibfield  {author} {\bibinfo {author} {\bibfnamefont {A.~F.}\ \bibnamefont
  {Falk}}, \bibinfo {author} {\bibfnamefont {H.}~\bibnamefont {Georgi}},
  \bibinfo {author} {\bibfnamefont {B.}~\bibnamefont {Grinstein}}, \ and\
  \bibinfo {author} {\bibfnamefont {M.~B.}\ \bibnamefont {Wise}},\ }\href
  {\doibase 10.1016/0550-3213(90)90591-Z} {\bibfield  {journal} {\bibinfo
  {journal} {Nucl.\ Phys.}\ }\textbf {\bibinfo {volume} {B343}},\ \bibinfo
  {pages} {1} (\bibinfo {year} {1990})}\BibitemShut {NoStop}%
\bibitem [{\citenamefont {Kiselev}()}]{Kiselev:2002vz}%
  \BibitemOpen
  \bibfield  {author} {\bibinfo {author} {\bibfnamefont {V.~V.}\ \bibnamefont
  {Kiselev}},\ }\href@noop {} {\ }\Eprint {http://arxiv.org/abs/hep-ph/0211021}
  {arXiv:hep-ph/0211021 [hep-ph]} \BibitemShut {NoStop}%
\bibitem [{\citenamefont {Bailey}\ \emph {et~al.}(2014)\citenamefont {Bailey}
  \emph {et~al.}}]{Bailey:2014tva}%
  \BibitemOpen
  \bibfield  {author} {\bibinfo {author} {\bibfnamefont {J.~A.}\ \bibnamefont
  {Bailey}} \emph {et~al.} (\bibinfo {collaboration} {Fermilab Lattice and MILC
  Collaborations}),\ }\href {\doibase 10.1103/PhysRevD.89.114504} {\bibfield
  {journal} {\bibinfo  {journal} {Phys.\ Rev.}\ }\textbf {\bibinfo {volume}
  {D89}},\ \bibinfo {pages} {114504} (\bibinfo {year} {2014})},\ \Eprint
  {http://arxiv.org/abs/1403.0635} {arXiv:1403.0635 [hep-lat]} \BibitemShut
  {NoStop}%
\bibitem [{\citenamefont {Harrison}\ \emph {et~al.}(2018)\citenamefont
  {Harrison}, \citenamefont {Davies},\ and\ \citenamefont
  {Wingate}}]{Harrison:2017fmw}%
  \BibitemOpen
  \bibfield  {author} {\bibinfo {author} {\bibfnamefont {J.}~\bibnamefont
  {Harrison}}, \bibinfo {author} {\bibfnamefont {C.}~\bibnamefont {Davies}}, \
  and\ \bibinfo {author} {\bibfnamefont {M.}~\bibnamefont {Wingate}} (\bibinfo
  {collaboration} {HPQCD Collaboration}),\ }\href {\doibase
  10.1103/PhysRevD.97.054502} {\bibfield  {journal} {\bibinfo  {journal}
  {Phys.\ Rev.}\ }\textbf {\bibinfo {volume} {D97}},\ \bibinfo {pages} {054502}
  (\bibinfo {year} {2018})},\ \Eprint {http://arxiv.org/abs/1711.11013}
  {arXiv:1711.11013 [hep-lat]} \BibitemShut {NoStop}%
\bibitem [{\citenamefont {McLean}\ \emph
  {et~al.}(2019{\natexlab{b}})\citenamefont {McLean}, \citenamefont {Davies},
  \citenamefont {Lytle},\ and\ \citenamefont {Koponen}}]{McLean:2019sds}%
  \BibitemOpen
  \bibfield  {author} {\bibinfo {author} {\bibfnamefont {E.}~\bibnamefont
  {McLean}}, \bibinfo {author} {\bibfnamefont {C.~T.~H.}\ \bibnamefont
  {Davies}}, \bibinfo {author} {\bibfnamefont {A.~T.}\ \bibnamefont {Lytle}}, \
  and\ \bibinfo {author} {\bibfnamefont {J.}~\bibnamefont {Koponen}},\ }\href
  {\doibase 10.1103/PhysRevD.99.114512} {\bibfield  {journal} {\bibinfo
  {journal} {Phys.\ Rev.}\ }\textbf {\bibinfo {volume} {D99}},\ \bibinfo
  {pages} {114512} (\bibinfo {year} {2019}{\natexlab{b}})},\ \Eprint
  {http://arxiv.org/abs/1904.02046} {arXiv:1904.02046 [hep-lat]} \BibitemShut
  {NoStop}%
\bibitem [{\citenamefont {Boyd}\ \emph
  {et~al.}(1995{\natexlab{b}})\citenamefont {Boyd}, \citenamefont {Grinstein},\
  and\ \citenamefont {Lebed}}]{Boyd:1994tt}%
  \BibitemOpen
  \bibfield  {author} {\bibinfo {author} {\bibfnamefont {C.~G.}\ \bibnamefont
  {Boyd}}, \bibinfo {author} {\bibfnamefont {B.}~\bibnamefont {Grinstein}}, \
  and\ \bibinfo {author} {\bibfnamefont {R.~F.}\ \bibnamefont {Lebed}},\ }\href
  {\doibase 10.1103/PhysRevLett.74.4603} {\bibfield  {journal} {\bibinfo
  {journal} {Phys.\ Rev.\ Lett.}\ }\textbf {\bibinfo {volume} {74}},\ \bibinfo
  {pages} {4603} (\bibinfo {year} {1995}{\natexlab{b}})},\ \Eprint
  {http://arxiv.org/abs/hep-ph/9412324} {arXiv:hep-ph/9412324 [hep-ph]}
  \BibitemShut {NoStop}%
\bibitem [{\citenamefont {Boyd}\ and\ \citenamefont
  {Lebed}(1997)}]{Boyd:1995tg}%
  \BibitemOpen
  \bibfield  {author} {\bibinfo {author} {\bibfnamefont {C.~G.}\ \bibnamefont
  {Boyd}}\ and\ \bibinfo {author} {\bibfnamefont {R.~F.}\ \bibnamefont
  {Lebed}},\ }\href {\doibase 10.1016/S0550-3213(96)00614-1} {\bibfield
  {journal} {\bibinfo  {journal} {Nucl.\ Phys.}\ }\textbf {\bibinfo {volume}
  {B485}},\ \bibinfo {pages} {275} (\bibinfo {year} {1997})},\ \Eprint
  {http://arxiv.org/abs/hep-ph/9512363} {arXiv:hep-ph/9512363 [hep-ph]}
  \BibitemShut {NoStop}%
\bibitem [{\citenamefont {Boyd}\ \emph {et~al.}(1996)\citenamefont {Boyd},
  \citenamefont {Grinstein},\ and\ \citenamefont {Lebed}}]{Boyd:1995sq}%
  \BibitemOpen
  \bibfield  {author} {\bibinfo {author} {\bibfnamefont {C.~G.}\ \bibnamefont
  {Boyd}}, \bibinfo {author} {\bibfnamefont {B.}~\bibnamefont {Grinstein}}, \
  and\ \bibinfo {author} {\bibfnamefont {R.~F.}\ \bibnamefont {Lebed}},\ }\href
  {\doibase 10.1016/0550-3213(95)00653-2} {\bibfield  {journal} {\bibinfo
  {journal} {Nucl.\ Phys.}\ }\textbf {\bibinfo {volume} {B461}},\ \bibinfo
  {pages} {493} (\bibinfo {year} {1996})},\ \Eprint
  {http://arxiv.org/abs/hep-ph/9508211} {arXiv:hep-ph/9508211 [hep-ph]}
  \BibitemShut {NoStop}%
\bibitem [{\citenamefont {Generalis}(1990)}]{Generalis:1990id}%
  \BibitemOpen
  \bibfield  {author} {\bibinfo {author} {\bibfnamefont {S.~C.}\ \bibnamefont
  {Generalis}},\ }\href {\doibase 10.1088/0954-3899/16/6/002} {\bibfield
  {journal} {\bibinfo  {journal} {J. Phys.}\ }\textbf {\bibinfo {volume}
  {G16}},\ \bibinfo {pages} {785} (\bibinfo {year} {1990})}\BibitemShut
  {NoStop}%
\bibitem [{\citenamefont {Reinders}\ \emph {et~al.}(1980)\citenamefont
  {Reinders}, \citenamefont {Rubinstein},\ and\ \citenamefont
  {Yazaki}}]{Reinders:1980wk}%
  \BibitemOpen
  \bibfield  {author} {\bibinfo {author} {\bibfnamefont {L.~J.}\ \bibnamefont
  {Reinders}}, \bibinfo {author} {\bibfnamefont {H.~R.}\ \bibnamefont
  {Rubinstein}}, \ and\ \bibinfo {author} {\bibfnamefont {S.}~\bibnamefont
  {Yazaki}},\ }\href {\doibase 10.1016/0370-2693(80)90596-1} {\bibfield
  {journal} {\bibinfo  {journal} {Phys.\ Lett.}\ }\textbf {\bibinfo {volume}
  {97B}},\ \bibinfo {pages} {257} (\bibinfo {year} {1980})},\ \bibinfo {note}
  {[Erratum: Phys.\ Lett.\ {\bf 100B}, 519 (1981)]}\BibitemShut {NoStop}%
\bibitem [{\citenamefont {Reinders}\ \emph {et~al.}(1981)\citenamefont
  {Reinders}, \citenamefont {Yazaki},\ and\ \citenamefont
  {Rubinstein}}]{Reinders:1981sy}%
  \BibitemOpen
  \bibfield  {author} {\bibinfo {author} {\bibfnamefont {L.~J.}\ \bibnamefont
  {Reinders}}, \bibinfo {author} {\bibfnamefont {S.}~\bibnamefont {Yazaki}}, \
  and\ \bibinfo {author} {\bibfnamefont {H.~R.}\ \bibnamefont {Rubinstein}},\
  }\href {\doibase 10.1016/0370-2693(81)90194-5} {\bibfield  {journal}
  {\bibinfo  {journal} {Phys.\ Lett.}\ }\textbf {\bibinfo {volume} {103B}},\
  \bibinfo {pages} {63} (\bibinfo {year} {1981})}\BibitemShut {NoStop}%
\bibitem [{\citenamefont {Reinders}\ \emph {et~al.}(1985)\citenamefont
  {Reinders}, \citenamefont {Rubinstein},\ and\ \citenamefont
  {Yazaki}}]{Reinders:1984sr}%
  \BibitemOpen
  \bibfield  {author} {\bibinfo {author} {\bibfnamefont {L.~J.}\ \bibnamefont
  {Reinders}}, \bibinfo {author} {\bibfnamefont {H.}~\bibnamefont
  {Rubinstein}}, \ and\ \bibinfo {author} {\bibfnamefont {S.}~\bibnamefont
  {Yazaki}},\ }\href {\doibase 10.1016/0370-1573(85)90065-1} {\bibfield
  {journal} {\bibinfo  {journal} {Phys.\ Rept.}\ }\textbf {\bibinfo {volume}
  {127}},\ \bibinfo {pages} {1} (\bibinfo {year} {1985})}\BibitemShut {NoStop}%
\bibitem [{\citenamefont {Djouadi}\ and\ \citenamefont
  {Gambino}(1994)}]{Djouadi:1993ss}%
  \BibitemOpen
  \bibfield  {author} {\bibinfo {author} {\bibfnamefont {A.}~\bibnamefont
  {Djouadi}}\ and\ \bibinfo {author} {\bibfnamefont {P.}~\bibnamefont
  {Gambino}},\ }\href {\doibase 10.1103/PhysRevD.49.3499,
  10.1103/PhysRevD.53.4111} {\bibfield  {journal} {\bibinfo  {journal} {Phys.\
  Rev.}\ }\textbf {\bibinfo {volume} {D49}},\ \bibinfo {pages} {3499} (\bibinfo
  {year} {1994})},\ \bibinfo {note} {[Erratum: Phys.\ Rev.\ {\bf D53}, 4111
  (1996)]},\ \Eprint {http://arxiv.org/abs/hep-ph/9309298}
  {arXiv:hep-ph/9309298 [hep-ph]} \BibitemShut {NoStop}%
\bibitem [{\citenamefont {Aaij}\ \emph
  {et~al.}(2017{\natexlab{a}})\citenamefont {Aaij} \emph
  {et~al.}}]{Aaij:2016qlz}%
  \BibitemOpen
  \bibfield  {author} {\bibinfo {author} {\bibfnamefont {R.}~\bibnamefont
  {Aaij}} \emph {et~al.} (\bibinfo {collaboration} {LHCb Collaboration}),\
  }\href {\doibase 10.1103/PhysRevD.95.032005} {\bibfield  {journal} {\bibinfo
  {journal} {Phys.\ Rev.}\ }\textbf {\bibinfo {volume} {D95}},\ \bibinfo
  {pages} {032005} (\bibinfo {year} {2017}{\natexlab{a}})},\ \Eprint
  {http://arxiv.org/abs/1612.07421} {arXiv:1612.07421 [hep-ex]} \BibitemShut
  {NoStop}%
\bibitem [{\citenamefont {Sirunyan}\ \emph {et~al.}(2019)\citenamefont
  {Sirunyan} \emph {et~al.}}]{Sirunyan:2019osb}%
  \BibitemOpen
  \bibfield  {author} {\bibinfo {author} {\bibfnamefont {A.~M.}\ \bibnamefont
  {Sirunyan}} \emph {et~al.} (\bibinfo {collaboration} {CMS Collaboration}),\
  }\href {\doibase 10.1103/PhysRevLett.122.132001} {\bibfield  {journal}
  {\bibinfo  {journal} {Phys.\ Rev.\ Lett.}\ }\textbf {\bibinfo {volume}
  {122}},\ \bibinfo {pages} {132001} (\bibinfo {year} {2019})},\ \Eprint
  {http://arxiv.org/abs/1902.00571} {arXiv:1902.00571 [hep-ex]} \BibitemShut
  {NoStop}%
\bibitem [{\citenamefont {Eichten}\ and\ \citenamefont
  {Quigg}(2019)}]{Eichten:2019gig}%
  \BibitemOpen
  \bibfield  {author} {\bibinfo {author} {\bibfnamefont {E.~J.}\ \bibnamefont
  {Eichten}}\ and\ \bibinfo {author} {\bibfnamefont {C.}~\bibnamefont
  {Quigg}},\ }\href {\doibase 10.1103/PhysRevD.99.054025} {\bibfield  {journal}
  {\bibinfo  {journal} {Phys.\ Rev.}\ }\textbf {\bibinfo {volume} {D99}},\
  \bibinfo {pages} {054025} (\bibinfo {year} {2019})},\ \Eprint
  {http://arxiv.org/abs/1902.09735} {arXiv:1902.09735 [hep-ph]} \BibitemShut
  {NoStop}%
\bibitem [{\citenamefont {Aubert}\ \emph {et~al.}(2010)\citenamefont {Aubert}
  \emph {et~al.}}]{Aubert:2009ac}%
  \BibitemOpen
  \bibfield  {author} {\bibinfo {author} {\bibfnamefont {B.}~\bibnamefont
  {Aubert}} \emph {et~al.} (\bibinfo {collaboration} {BaBar Collaboration}),\
  }\href {\doibase 10.1103/PhysRevLett.104.011802} {\bibfield  {journal}
  {\bibinfo  {journal} {Phys.\ Rev.\ Lett.}\ }\textbf {\bibinfo {volume}
  {104}},\ \bibinfo {pages} {011802} (\bibinfo {year} {2010})},\ \Eprint
  {http://arxiv.org/abs/0904.4063} {arXiv:0904.4063 [hep-ex]} \BibitemShut
  {NoStop}%
\bibitem [{\citenamefont {Glattauer}\ \emph {et~al.}(2016)\citenamefont
  {Glattauer} \emph {et~al.}}]{Glattauer:2015teq}%
  \BibitemOpen
  \bibfield  {author} {\bibinfo {author} {\bibfnamefont {R.}~\bibnamefont
  {Glattauer}} \emph {et~al.} (\bibinfo {collaboration} {Belle
  Collaboration}),\ }\href {\doibase 10.1103/PhysRevD.93.032006} {\bibfield
  {journal} {\bibinfo  {journal} {Phys.\ Rev.}\ }\textbf {\bibinfo {volume}
  {D93}},\ \bibinfo {pages} {032006} (\bibinfo {year} {2016})},\ \Eprint
  {http://arxiv.org/abs/1510.03657} {arXiv:1510.03657 [hep-ex]} \BibitemShut
  {NoStop}%
\bibitem [{\citenamefont {Lees}\ \emph {et~al.}(2019)\citenamefont {Lees} \emph
  {et~al.}}]{Dey:2019bgc}%
  \BibitemOpen
  \bibfield  {author} {\bibinfo {author} {\bibfnamefont {J.~P.}\ \bibnamefont
  {Lees}} \emph {et~al.} (\bibinfo {collaboration} {BaBar Collaboration}),\
  }\href {\doibase 10.1103/PhysRevLett.123.091801} {\bibfield  {journal}
  {\bibinfo  {journal} {Phys.\ Rev.\ Lett.}\ }\textbf {\bibinfo {volume}
  {123}},\ \bibinfo {pages} {091801} (\bibinfo {year} {2019})},\ \Eprint
  {http://arxiv.org/abs/1903.10002} {arXiv:1903.10002 [hep-ex]} \BibitemShut
  {NoStop}%
\bibitem [{\citenamefont {Le~Yaouanc}\ \emph {et~al.}(2009)\citenamefont
  {Le~Yaouanc}, \citenamefont {Oliver},\ and\ \citenamefont
  {Raynal}}]{LeYaouanc:2008pq}%
  \BibitemOpen
  \bibfield  {author} {\bibinfo {author} {\bibfnamefont {A.}~\bibnamefont
  {Le~Yaouanc}}, \bibinfo {author} {\bibfnamefont {L.}~\bibnamefont {Oliver}},
  \ and\ \bibinfo {author} {\bibfnamefont {J.-C.}\ \bibnamefont {Raynal}},\
  }\href {\doibase 10.1103/PhysRevD.79.014023} {\bibfield  {journal} {\bibinfo
  {journal} {Phys.\ Rev.}\ }\textbf {\bibinfo {volume} {D79}},\ \bibinfo
  {pages} {014023} (\bibinfo {year} {2009})},\ \Eprint
  {http://arxiv.org/abs/0808.2983} {arXiv:0808.2983 [hep-ph]} \BibitemShut
  {NoStop}%
\bibitem [{\citenamefont {Aaij}\ \emph
  {et~al.}(2017{\natexlab{b}})\citenamefont {Aaij} \emph
  {et~al.}}]{Aaij:2017svr}%
  \BibitemOpen
  \bibfield  {author} {\bibinfo {author} {\bibfnamefont {R.}~\bibnamefont
  {Aaij}} \emph {et~al.} (\bibinfo {collaboration} {LHCb Collaboration}),\
  }\href {\doibase 10.1103/PhysRevD.96.112005} {\bibfield  {journal} {\bibinfo
  {journal} {Phys.\ Rev.}\ }\textbf {\bibinfo {volume} {D96}},\ \bibinfo
  {pages} {112005} (\bibinfo {year} {2017}{\natexlab{b}})},\ \Eprint
  {http://arxiv.org/abs/1709.01920} {arXiv:1709.01920 [hep-ex]} \BibitemShut
  {NoStop}%
\bibitem [{\citenamefont {Abdesselam}\ \emph
  {et~al.}(2019{\natexlab{b}})\citenamefont {Abdesselam} \emph
  {et~al.}}]{Abdesselam:2019wbt}%
  \BibitemOpen
  \bibfield  {author} {\bibinfo {author} {\bibfnamefont {A.}~\bibnamefont
  {Abdesselam}} \emph {et~al.} (\bibinfo {collaboration} {Belle
  Collaboration}),\ }in\ \href@noop {} {\emph {\bibinfo {booktitle} {{10th
  International Workshop on the CKM Unitarity Triangle (CKM 2018) Heidelberg,
  Germany, September 17-21, 2018}}}}\ (\bibinfo {year} {2019})\ \Eprint
  {http://arxiv.org/abs/1903.03102} {arXiv:1903.03102 [hep-ex]} \BibitemShut
  {NoStop}%
\bibitem [{\citenamefont {Huang}\ \emph {et~al.}(2018)\citenamefont {Huang},
  \citenamefont {Li}, \citenamefont {Lu}, \citenamefont {Paracha},\ and\
  \citenamefont {Wang}}]{Huang:2018nnq}%
  \BibitemOpen
  \bibfield  {author} {\bibinfo {author} {\bibfnamefont {Z.-R.}\ \bibnamefont
  {Huang}}, \bibinfo {author} {\bibfnamefont {Y.}~\bibnamefont {Li}}, \bibinfo
  {author} {\bibfnamefont {C.-D.}\ \bibnamefont {Lu}}, \bibinfo {author}
  {\bibfnamefont {M.~A.}\ \bibnamefont {Paracha}}, \ and\ \bibinfo {author}
  {\bibfnamefont {C.}~\bibnamefont {Wang}},\ }\href {\doibase
  10.1103/PhysRevD.98.095018} {\bibfield  {journal} {\bibinfo  {journal}
  {Phys.\ Rev.}\ }\textbf {\bibinfo {volume} {D98}},\ \bibinfo {pages} {095018}
  (\bibinfo {year} {2018})},\ \Eprint {http://arxiv.org/abs/1808.03565}
  {arXiv:1808.03565 [hep-ph]} \BibitemShut {NoStop}%
\bibitem [{\citenamefont {Vaquero}\ \emph {et~al.}(2019)\citenamefont
  {Vaquero}, \citenamefont {DeTar}, \citenamefont {El-Khadra}, \citenamefont
  {Kronfeld}, \citenamefont {Laiho},\ and\ \citenamefont {Van~de
  Water}}]{Vaquero:2019ary}%
  \BibitemOpen
  \bibfield  {author} {\bibinfo {author} {\bibfnamefont {A.}~\bibnamefont
  {Vaquero}}, \bibinfo {author} {\bibfnamefont {C.}~\bibnamefont {DeTar}},
  \bibinfo {author} {\bibfnamefont {A.~X.}\ \bibnamefont {El-Khadra}}, \bibinfo
  {author} {\bibfnamefont {A.~S.}\ \bibnamefont {Kronfeld}}, \bibinfo {author}
  {\bibfnamefont {J.}~\bibnamefont {Laiho}}, \ and\ \bibinfo {author}
  {\bibfnamefont {R.~S.}\ \bibnamefont {Van~de Water}},\ }in\ \href@noop {}
  {\emph {\bibinfo {booktitle} {{17th Conference on Flavor Physics and CP
  Violation (FPCP 2019) Victoria, BC, Canada, May 6-10, 2019}}}}\ (\bibinfo
  {year} {2019})\ \Eprint {http://arxiv.org/abs/1906.01019} {arXiv:1906.01019
  [hep-lat]} \BibitemShut {NoStop}%
\end{thebibliography}%
\end{document}